\documentclass[useAMS,usenatbib,fleqn]{mn2e}

\usepackage{graphicx}
\usepackage{amsmath}
\usepackage{amssymb}  
\usepackage{url}
\usepackage{times}

\def \aap{A\&A}

\def \apj{ApJ}
\def \apjl{ApJ}
\def \apjs{ApJS}
\def \araa{ARA\&A}

\def \mnras{MNRAS}

\def \nat{Nat}

\def \pasp{PASP}

\def \rmxaa{RMXAA}

\newcommand{\NH}{\ensuremath{{\it N}_\mathrm{H}}}
\newcommand{\NHI}{\ensuremath{{\it N}_\mathrm{HI}}}
\newcommand{\NOVI}{\ensuremath{{\it N}_\mathrm{OVI}}}

\newcommand{\Ha}{\ensuremath{\rm H\alpha}}
\newcommand{\lya}{\ensuremath{\rm Ly\alpha}}

\newcommand{\kms}{\ensuremath{\rm km\,s^{-1}}}
\newcommand{\kmsMpc}{\ensuremath{\rm km\,s^{-1}\,Mpc^{-1}}}
\newcommand{\cmm}{\rm cm\ensuremath{^{-2}}}
\newcommand{\cmmm}{\rm cm\ensuremath{^{-3}}}
\newcommand{\msun}{\ensuremath{{\it M}_\odot}}
\newcommand{\msunyr}{\ensuremath{{\it M}_{\odot}\,{\rm yr}^{-1}}}
\newcommand{\msunyrkpcc}{\ensuremath{{\it M}_{\odot}\,{\rm yr}^{-1}\,{\rm kpc}^{-2}}}

\newcommand{\ergscmm}{\ensuremath{{\rm erg~s}^{-1}~{\rm cm}^{-2}}}

\newcommand{\AlII}{\hbox{{\rm Al}{\sc \,ii}}}
\newcommand{\AlIII}{\hbox{{\rm Al}{\sc \,iii}}}
\newcommand{\CII}{\hbox{{\rm C}{\sc \,ii}}}

\newcommand{\CIII}{\hbox{{\rm C}{\sc \,iii}}}
\newcommand{\CIV}{\hbox{{\rm C}{\sc \,iv}}}

\newcommand{\FeIII}{\hbox{{\rm Fe}{\sc \,iii}}}
\newcommand{\HeII}{\hbox{{\rm He}{\sc \,ii}}}
\newcommand{\HI}{\hbox{{\rm H}{\sc \,i}}}
\newcommand{\HII}{\hbox{{\rm H}{\sc \,ii}}}

\newcommand{\MgII}{\hbox{{\rm Mg}{\sc \,ii}}}

\newcommand{\NII}{\hbox{{\rm N}{\sc \,ii}}}
\newcommand{\NV}{\hbox{{\rm N}{\sc \,v}}}
\newcommand{\OI}{O{\sc\,i}}

\newcommand{\OVI}{{\rm O}{\sc \,vi}}

\newcommand{\SiII}{\hbox{{\rm Si}{\sc \,ii}}}
\newcommand{\SiIII}{\hbox{{\rm Si}{\sc \,iii}}}
\newcommand{\SiIV}{\hbox{{\rm Si}{\sc \,iv}}}

\newcommand{\nH}{\ensuremath{{\it n}_{\rm H}}}

\title[Enriched gas 50~kpc from a faint $z=2.466$
  galaxy]{Metal-enriched, sub-kiloparsec gas clumps in the
  circumgalactic medium of a faint $\bmath{z=2.5}$ galaxy\thanks{Data
    and code used for this paper are available at
    \protect\url{https://github.com/nhmc/LAE}; doi:10.5281/zenodo.12321}}

\author[N. Crighton et al.]{Neil
  H. M. Crighton$^1$\thanks{neilcrighton@gmail.com}, Joseph
F. Hennawi$^2$, Robert A. Simcoe$^3$, Kathy L. Cooksey$^4$, \newauthor Michael
T. Murphy$^1$, Michele Fumagalli$^{5,6}$, J. Xavier Prochaska$^7$ and
Tom Shanks$^5$
\\
\\
$^1$ Centre for Astrophysics and Supercomputing, Swinburne
  University of Technology, Hawthorn, Victoria 3122, Australia. \\
$^2$ Max-Planck-Institut f\"ur Astronomie, K\"onigstuhl 17, 69117
  Heidelberg, Germany. \\
$^3$ MIT-Kavli Center for Astrophysics and Space Research, 77
  Massachusetts Avenue, \#37–6640, Cambridge, MA 02139, USA. \\
$^4$ University of Hawai`i at Hilo, 200 W. K\=awili St., Hilo,
  HI 96720, USA. \\
$^5$ Institute for Computational Cosmology, Department of
  Physics, Durham University, South Road, Durham, DH1 3LE, UK. \\
$^6$ Carnegie Observatories, 813 Santa Barbara Street, Pasadena,
  CA 91101, USA. \\
$^7$ Department of Astronomy and
  Astrophysics, UCO/Lick Observatory; University of California, 1156
  High Street, Santa Cruz, CA 95064, USA.
}

\begin{document}

\date{Accepted xxxx. Received xxxx; in original form xxxx}

\pagerange{\pageref{firstpage}--\pageref{lastpage}} \pubyear{xxxx}

\maketitle

\label{firstpage}

\begin{abstract}

We report the serendipitous detection of a 0.2 L$^*$, \lya\ emitting
galaxy at redshift 2.5 at an impact parameter of 50~kpc from a bright
background QSO sightline. A high-resolution spectrum of the QSO
reveals a partial Lyman-limit absorption system
($\NHI=10^{16.94\pm0.10}$~\cmm) with many associated metal absorption
lines at the same redshift as the foreground galaxy. Using
photoionization models that carefully treat measurement errors and
marginalise over uncertainties in the shape and normalisation of the
ionizing radiation spectrum, we derive the total hydrogen column
density $\NH=10^{19.4\pm0.3}~\cmm$, and show that all the absorbing
clouds are metal enriched, with $Z=0.1$--$0.6~Z_\odot$. These
metallicities and the system's large velocity width ($436$~\kms)
suggest the gas is produced by an outflowing wind. Using an expanding
shell model we estimate a mass outflow rate of $\sim5\msunyr$. Our
photoionization model yields extremely small sizes ($<$100--500 pc)
for the absorbing clouds, which we argue is typical of high column
density absorbers in the circumgalactic medium (CGM). Given these
small sizes and extreme kinematics, it is unclear how the clumps
survive in the CGM without being destroyed by hydrodynamic
instabilities. The small cloud sizes imply that even state-of-the-art
cosmological simulations require more than a $1000$-fold improvement
in mass resolution to resolve the hydrodynamics relevant for cool gas
in the CGM.

\end{abstract}


\section{Introduction}

Observations of the gaseous halos surrounding galaxies---the
circumgalactic medium (CGM)---allow us to constrain two of the most
poorly understood aspects of galaxy formation: galactic-scale winds
and gas accretion onto galaxies. Detecting this gas in emission at
high redshift is possible, but only in extreme environments or with
very deep observations
\citep[e.g.][]{Steidel11,Rauch13,Cantalupo14,Martin14} using current
facilities. The diffuse gas comprising galactic-scale winds and
accreting gas can be relatively easily measured as rest-frame UV
absorption features however, which are imprinted on the spectrum of
the galaxy itself (`down-the-barrel'), or on a background QSO at a
small impact parameter from the galaxy.

Observations of blue-shifted absorption in galaxy spectra have shown
that galactic-scale winds are common from $z\sim0.5$
\citep{Weiner09,Martin12,Rubin14} to $z\sim3$
\citep{Pettini01_LBG,Adelberger05,Bielby11,Steidel10}. In some cases
redshifted absorption is also seen, suggesting the presence of
metal-enriched, inflowing gas \citep{Rubin12,Martin12}. However, the
faintness of the background galaxies used by these studies mean that
only low resolution spectra can be used, and the absorption lines are
not resolved. Therefore the metallicity, ionization state, and volume
density of the gas remain poorly determined. Background QSOs are much
brighter than galaxies, and thus a high-resolution spectrum of a QSO
at small impact parameter ($\lesssim100$~kpc) from a foreground galaxy
can resolve individual metal transitions in the galaxy's CGM. Precise
column density measurements then enable us to tightly constrain the
physical properties of the gas using photoionization models.

A growing sample of QSO absorber--galaxy pairs is being assembled by
searching for galaxies around strong $z\sim3$ damped-\lya\ systems
$\NHI > 10^{20.3} {\rm cm^{-2}}$
\citep[e.g.][]{Djorgovski96_DLA,Moller02,Peroux11_SinfoniI,Fynbo13}. This
technique has found one example that may be caused by accreting gas
\citep{Bouche13} and others that may be produced by outflowing gas
\citep[e.g.][]{Noterdaeme12,Krogager13,Peroux13_SinfoniIV}.  However,
a drawback of this approach is that by construction, these pairs only
represent the strongest, often metal-rich absorbers. In addition, the
physical properties of individual absorbing components cannot be
measured, as line saturation and the \lya\ damping wings make it
difficult to divide the total observed \NHI\ between different metal
components.

An alternative approach is to search for galaxies close to QSO
sightlines without any absorption pre-selection. This allows a census
of gas around galaxies for a wide range of absorption
properties. Several surveys have been undertaken to assemble such
samples
\citep{Adelberger03,Adelberger05,Crighton11,Rudie12}. \citet{Simcoe06}
performed the first photoionization modelling of a `partial' Lyman
limit system (LLS)\footnote{We define a partial Lyman limit system as
  having an optical depth $<1$ at the Lyman limit.}  with
$\NHI\approx10^{16}$~\cmm\ at an impact parameter of $115$~kpc from a
$z=2.3$ galaxy found by one of these surveys. The absorption they see
can be explained by small, $\sim 100$~pc-scale, metal enriched gas
clumps suggestive of an outflowing wind.  Recently
\citet{Crighton13_cma} analysed a different galaxy-absorber pair at
$z=2.4$ with an impact parameter of $55$~kpc. They detected low
metallicity gas with properties consistent with those expected for
`cold-mode' accretion. They also detect metal enriched gas in the same
absorber. However, they found that the inferred metallicity of the
higher-$Z$ gas depends strongly on the shape assumed for the ionizing
spectrum used when modelling the clouds.
 
In this work we report the discovery of a partial Lyman-limit system
with $\NHI = 10^{16.9}~\cmm$ caused by metal-enriched gas that is
coincident in redshift with a \lya-emitting galaxy (LAE), fainter than
those selected with the traditional Lyman break techniques. The
absorber is at an impact parameter $\rho=50$~kpc from the galaxy. We
use a new modelling procedure which marginalises over uncertainties in
the ionizing radiation spectrum to make robust measurements of the
metallicity and volume density of the gas.

Our paper is structured as follows. In Section~\ref{s_gal} we describe
the galaxy properties. Section~\ref{s_abs} describes the properties of
the partial LLS, and how we perform \textsc{cloudy} modelling to
derive the metallicity and density of the absorbing
gas. Section~\ref{s_models} considers different scenarios that
reproduce our observations and discusses implications for simulations
of the CGM. Section~\ref{s_sim} discusses the implications of our
results for simulations of the CGM. Section~\ref{s_summary} summarises
our results. The appendices give details about our photoionization
modelling method. We use a Planck 2013 cosmology ($\Omega_{\rm
  M}=0.31$, $\Omega_\Lambda=0.69$, $H_0 =68~\kmsMpc$, Planck
Collaboration \citeyear{Planck13_XVI}), and all distances listed are
proper (not comoving) unless stated otherwise.

\section{The \lya-emitting Faint Galaxy at \lowercase{$\mathbf{z=2.466}$}}
\label{s_gal}

\begin{figure*}
\includegraphics[width=\linewidth]{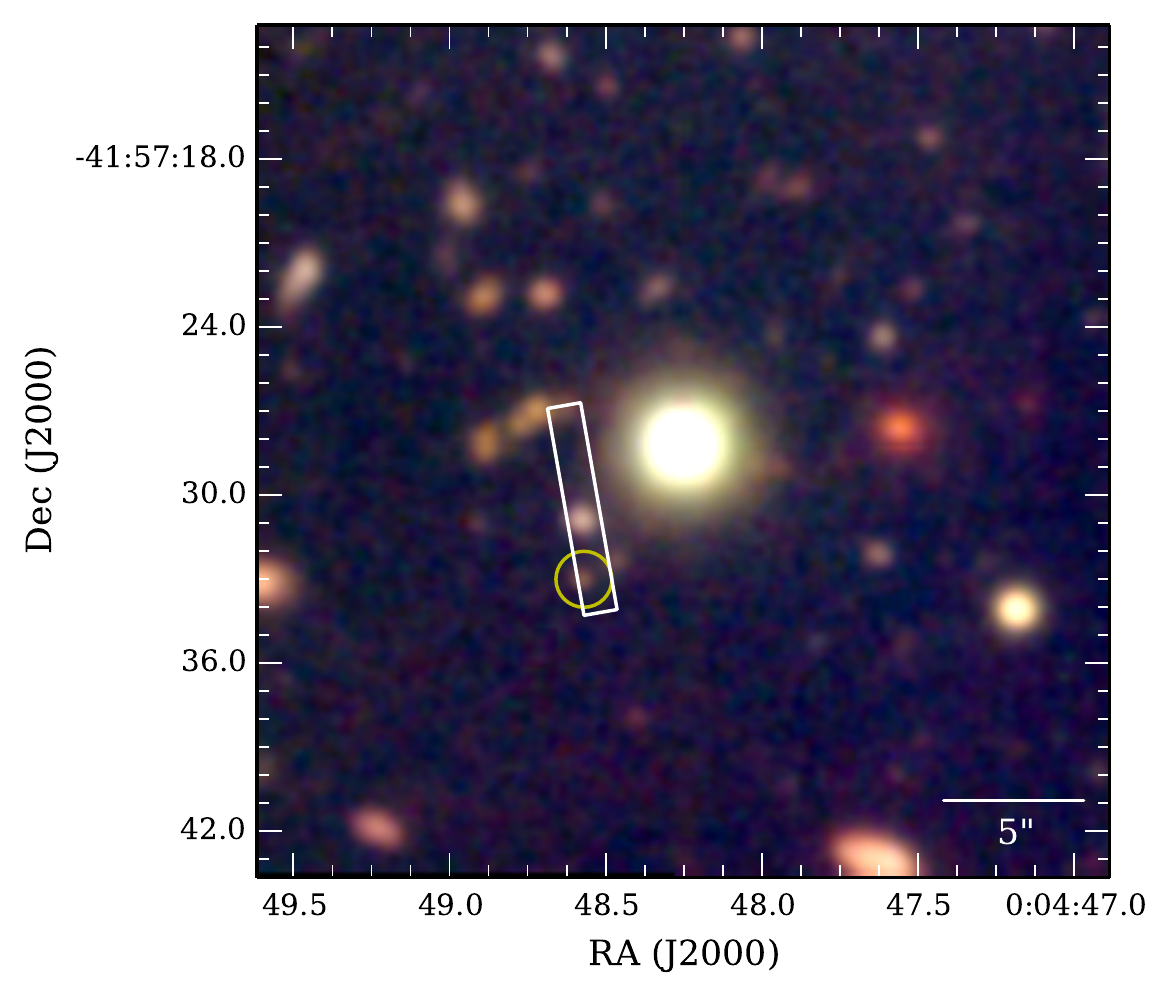}
\caption{\label{f_im} $u'$, $g'$ and $r'$ composite from
  Magellan/MagIC imaging showing the QSO (centre) and the nearby slit
  where \lya\ emission is seen at $z=2.466$. The circled object is the
  $r=25.9$ continuum source we associate with the \lya\ line. It is
  offset 3.53\arcsec\ east and 4.68\arcsec\ south of the QSO,
  corresponding to an impact parameter of 49 proper kpc at
  $z=2.466$. The object at the centre of the slit is an unrelated
  galaxy at $z < 2$.}
\end{figure*}

The galaxy was serendipitously discovered in a survey for $z\sim 2.5$
galaxies around QSO sightlines (Crighton et al. in preparation, Cooksey et
al. in preparation). The original aim of these observations was to confirm
galaxy candidates with $r'$ magnitude $< 25.5$ around the sightline to
the QSO Q0002$-$422 (RA 00h04m48.1s,
Dec. -41$^\circ$57$^\prime$29$^{\prime\prime}$ [J2000]
$z_\mathrm{QSO}=2.76$, $r=17.4$). This QSO was selected without
reference to any absorption in the spectrum (i.e. the presence of a
DLA or Lyman-limit system). The galaxy we analyse in this paper is too
faint to satisfy the $r'<25.5$ selection criterion, but happened to
fall inside a slit targeting a brighter object near the QSO.  It was
detected by its \lya\ emission line, which we then linked to a faint
continuum source in deep imaging of the field.  Fig.~\ref{f_im} shows
a false colour image generated from this deep imaging in the $u'$,
$g'$ and $r'$ bands. This imaging was taken with the MagIC CCD camera
on the Magellan Clay telescope in October 2005, with total exposure
times of 1.5 hours each in $g'$ and $r'$ and 5 hours in $u'$. The
images were bias-subtracted, flat-fielded, registered and combined
using standard IRAF procedures and smoothed to a common seeing scale
of 0.6\arcsec (set by $u'$) to facilitate measurement of isophotal
colours. The $5\sigma$ detection limits for a point source in the
combined images are 26.3 ($u'$), 27.7 ($g'$) and 28.0 ($r'$).

The QSO is shown at the centre of Fig.~\ref{f_im}, and the slit in
which the emitter falls is on left. The brighter object at the centre
of this slit was the galaxy candidate originally targeted. Its
redshift is uncertain, but the absence of \lya\ forest absorption
requires $z < 2$. We identify the fainter, circled object as the
continuum source associated with the emitter. Sextractor was used to
measure magnitudes for this source in a 1.5\arcsec\ aperture, giving
$u=26.8\pm0.2$, $g=26.08\pm0.08$ and $r=25.90 \pm 0.08$ (AB).  Its
colours satisfy the BX selection criteria, designed to select
star-forming galaxies in the redshift range $2.0<z<2.5$
\citep{Steidel04} based on their continuum emission.

\begin{figure}
\includegraphics[width=\linewidth]{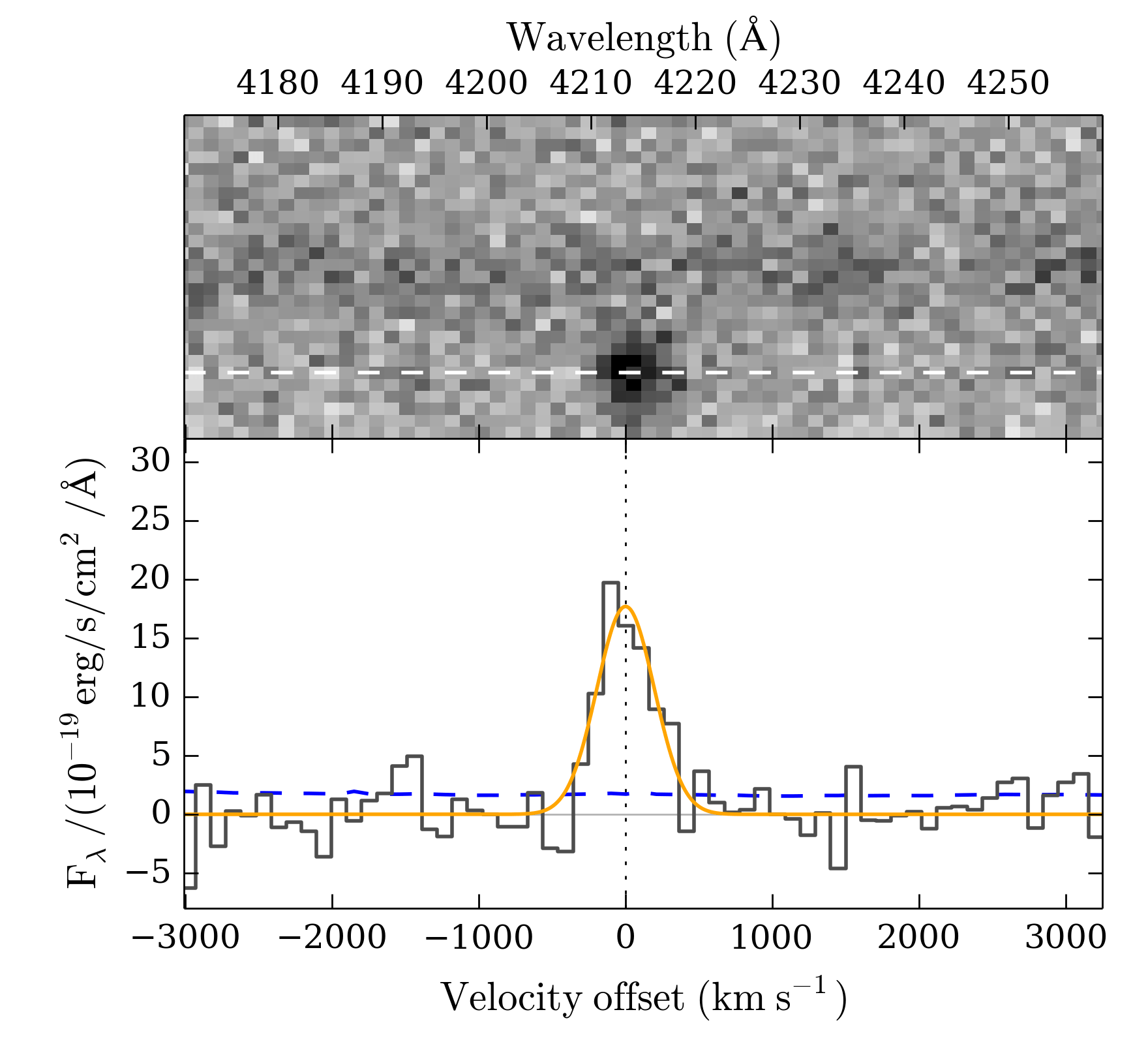}
\caption{\label{f_fors} FORS spectrum showing the \lya\ emission
  line. The top panel shows the emission line, offset from the trace
  of the central object that was the original target (which has
  $z<2$). A dashed white line shows the assumed trace used to extract
  the emission line. The bottom panel shows the extracted 1-d spectrum
  with the $1\sigma$ error array (blue dashed line), and a Gaussian
  profile fitted to the line, assuming a width given by the spectral
  resolution. The line is unresolved, implying an intrinsic width
  significantly less than 460~\kms, corresponding to the FORS spectral
  resolving power of 650. The velocity scale is relative to
  $z=2.466$.}
\end{figure}

We obtained the galaxy spectrum during program 091.A-0698 (PI
Crighton) using FORS2 at the Very Large Telescope (VLT) with the
\texttt{600B+22} grism and a 1.2\arcsec\ slit, resulting in a
resolution of 650 ($\sim 460~\kms$). A total of five 30 minute
exposures were taken over the 8th and 9th of September 2013 in clear
conditions with seeing $0.8-1\arcsec$. The exposures were
flat-fielded, combined and wavelength-calibrated using
Low-Redux\footnote{\url{http://www.ucolick.org/~xavier/LowRedux}}.
Observations of a standard star were used to flux calibrate the 1-d
spectrum, which covers a wavelength range $3600-6000$~\AA. The
\lya\ line that revealed the galaxy is shown in Fig.~\ref{f_fors}. It
is offset from the trace of the brighter, original target seen in the
centre of the top panel. By fitting a Gaussian to the line, we measure
an emission redshift $z_\mathrm{\lya}=2.4659\pm0.0003$. The line is
not resolved in this spectrum, so must have an intrinsic FWHM
significantly less than $460$~\kms. The FORS spectral resolution is
not high enough to separate the [OII] doublet ($\lambda3727$,
$\lambda3729$), which has a separation of $220$~\kms. However, there
are no other emission lines detected in the spectrum.  This means that
the line is unlikely to be caused by [OII] at $z=0.13$, because then
we would also expect to detect H-$\beta$ ($\lambda4863$) and [OIII]
($\lambda3727$, $\lambda3729$).
The linewidth is much narrower than expected for broad lines from a
QSO, and no \CIV\ emission at $z=2.4659$ is detected. Therefore the
emission is unlikely to be caused by an active galactic nucleus
(AGN). \lya\ is typically offset bluewards from the systemic redshift
in brighter Lyman-break selected galaxies by $300 \pm 125~\kms$
\citep{Rakic11}. For the small sample of fainter LAEs where \Ha\ has
been measured in addition to \lya, a smaller blue offset of
$220\pm30~\kms$ is found \citep{Yang11, Hashimoto13,
  Hashimoto_err13}. Therefore we adopt an intrinsic redshift
$z_\mathrm{gal} = z_\mathrm{\lya} -
(220\,\mathrm{\kms}/c)(1+z_\mathrm{\lya}) = 2.4636$ with an
uncertainty of $125~\kms$.

\subsection{The measured star-formation rate}

The \lya\ line flux is $1.1 \times 10^{-17}~\ergscmm$,
 which implies a lower limit to the star-formation rate SFR$_{\lya} >
 0.5\msunyr$ assuming case B recombination \citep[][section
   14.2.3]{Draine11} and no dust extinction. The rest-frame UV
 continuum gives an independent measure of the star-formation
 rate. Using the mean of the $g'$ and $r'$-band magnitude to estimate
 the rest-frame UV continuum magnitude $M_{1700}$, then applying a
 UV-to-SFR conversion factor of $7.2\times10^{-19}$ (appropriate for a
 Kroupa IMF and $\log_{10}[Z/Z_\odot]=-0.5$, \citealt{Madau14})
 results in a SFR$_\mathrm{UV}\approx1.5\msunyr$.
This is a factor of three larger than SFR$_{\lya}$. This discrepancy
can easily be explained by slit losses, which we expect to be large as
the was not centred on the slit, or preferential scattering or dust
absorption of \lya\ photons with respect to the continuum
emission. The rest-frame UV luminosity corresponds to $0.2 L^*$, using
$M_{1700}^*=-21.0$ \citep{Reddy09}.

Would this galaxy be classified as an LAE, as studied
by narrow band surveys \citep{Gawiser07, Nilsson11}?  LAEs are
empirically classified as satisfying rest equivalent width
($W_{r}\gtrsim20$ \AA\ in the rest frame) and line flux criteria
($\gtrsim 1\times10^{-17}~\ergscmm$). No continuum is detected in the
FORS spectrum, and conservatively adopting the $1 \sigma$ error level
as an upper limit to the continuum gives a lower limit $W_r > 15$
\AA. We can also estimate the continuum flux using the $g$ band
magnitude (the emission line does not make a significant contribution
to this band), which yields $W_r > 16 $~\AA. This is also a lower
limit as slit losses could be significant.  We conclude from the
$\lya$ line strength and equivalent width lower limits that this
object would indeed be selected as an LAE.

\subsection{Galaxy mass and SED fitting}

\begin{figure}
\includegraphics[width=\linewidth]{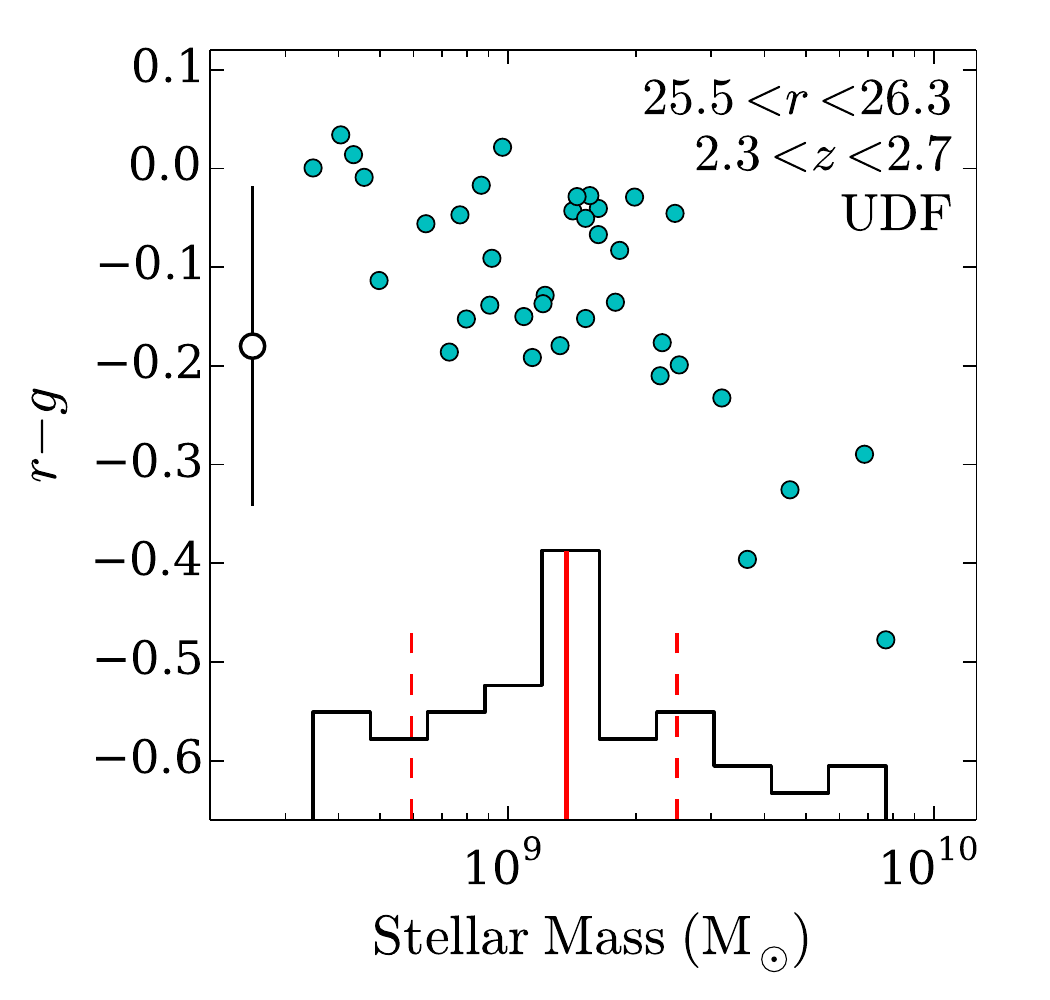}
\caption{\label{f_mass} $r-g$ colours as a function of stellar mass
  for galaxies in the Hubble Ultra Deep Field with a similar redshift
  ($2.3<z<2.7$) and magnitude ($25.5<r<26.3$) to the galaxy in this
  work. Stellar masses are taken from \citet{daCunha13}. The open
  circle shows $r - g$ and its $1\sigma$ uncertainty for our $z=2.5$
  galaxy (its $x$-axis position is arbitrary). The 16th, 50th (median)
  and 84th percentiles of the stellar mass are $10^{8.8}$, $10^{9.1}$
  and $10^{9.4}$\msun, shown by red vertical lines. This stellar mass
  range is lower than that of brighter BX-selected galaxies
  ($M_*\sim2\times10^{10}$\msun, \citealt{Shapley05}). }
\end{figure}

Currently only rest-frame UV magnitudes are available for the galaxy,
which makes it difficult to measure a stellar mass via spectral energy
distribution (SED) modelling.  However, the faint UV magnitudes
suggest a smaller stellar mass than is typical of BX-selected
galaxies. Fig.~\ref{f_mass} shows the distribution of $r-g$ colours
and stellar masses in the Hubble Ultra Deep Field (UDF) for galaxies
with a similar $r$ magnitude and redshift to the galaxy in this work
(from the SED fitting performed by \citealt{daCunha13}). The median
stellar mass of these objects is $10^{9.14}$\msun, with 16th and 84th
percentiles of $10^{8.8}$ and $10^{9.4}$\msun. This range is
significantly smaller than the typical stellar mass of BX-selected
galaxies with $R<25.5$,
$10^{10.32\pm0.51}$\msun\ \citep{Shapley05}. Moreover, we
demonstrated in the previous section that this galaxy would be
classified as an LAE. Clustering analyses show that LAEs have a typical
halo mass $\log_{10}({\rm M}/\msun)=10.9^{+0.5}_{-0.9}$
\citep{Gawiser07}, smaller than that of BX-selected galaxies (${\rm
  M}/\msun=10^{11.5-12.0}$,
\citealt{Adelberger05_cluster,Conroy08,Bielby13}). This lower halo
mass is consistent with the value found by converting a stellar mass
of $10^{9.14}$\msun\ to a halo mass using the relation from
\citet{Moster13}: $10^{11.4}$\msun.

Keeping the limitations of SED modelling in mind, we used the measured
$u'$, $g'$ and $r'$ colours to estimate the galaxy's stellar mass with
the SED fitting code \textsc{MAGPHYS} \citep{daCunha08}.  We corrected for
\HI\ absorption from the intergalactic medium (IGM) in the $u'$ and
$g'$ bands using \textsc{IGMtransmission}\footnote{Available for
  download from \url{http://code.google.com/p/igmtransmission}.}
\citep{Harrison11}, which calculates the mean absorption using the
transmission curves of \citet{Meiksin06}. The inferred stellar mass
(16th, 50th and 84th percentiles of $10^{8.8}$, $10^{9.4}$ and
$10^{9.7}$\msun) is consistent with the range given above for UDF
galaxies with a similar magnitude and redshift. These models also
predict a low SFR, similar to that estimated from the rest-frame UV
magnitude, and little dust extinction. A low dust extinction is
consistent with the low extinction measured for LAEs, $E(B-V) < 0.02$
\citep{Gawiser07}.
We conclude that it is likely the galaxy has a stellar mass of
$\sim10^{9.1}\msun$ (corresponding to a specific SFR
$\sim1.1$~Gyr$^{-1}$) and halo mass of $\sim10^{11.4}$\msun. We adopt
a fiducial halo mass of $10^{11.4}$\msun, which corresponds to a
virial radius of $60$~kpc, and indicate where our conclusions would
change assuming a higher halo mass. Deep rest-frame optical and
near-IR imaging are required for a more precise measurement of the
galaxy's stellar mass.

\section{The nearby partial Lyman-limit system}
\label{s_abs}

Quasar Q0002$-$422, initially identified by its strong \lya\ emission
in a prism survey \citep{Osmer76}, has been extensively observed with
the Ultraviolet and Visual Echelle Spectrograph (UVES) on the VLT in
Chile. We retrieved 49.5 hours of exposures from the ESO data archive,
spread over 43 individual exposures ranging between 3600 and 6500\,s
in duration each. These were observed under two different programs,
166.A-0106 (PI Bergeron) and 185.A-0745 (PI Molaro)\footnote{These
  programs observed this QSO due to its extreme luminosity and the
  presence of strong metal line systems at redshifts unrelated to the
  galaxy we study in this work.}, over 2001 July--2002 September and
2010 October--2012 November, during which the prevailing observing
conditions varied substantially, delivering seeing of 0.5--1.5 and
0.6--1.3 arcseconds, respectively. The slit-widths varied between the
two programs (1.0 and 0.8 arcseconds, respectively), so the spectral
resolutions obtained in the individual exposures varied between
$\sim45000$ and $\sim60000$.

The ESO UVES Common Pipeline Language data reduction software was used
to optimally extract and calibrate the relative quasar flux and
wavelength scales, and {\sc uves\_popler}\footnote{Written and
  maintained by MTM at
  \url{http://astronomy.swin.edu.au/~mmurphy/UVES\_popler}.} was used
to combine the many exposures into a single, normalised spectrum on a
vacuum-heliocentric wavelength scale. The 43 exposures were taken with
a wide variety of UVES wavelength settings which, when combined, cover
3050 to 9760\,\AA. The final signal-to-noise ratio (S/N) is 25 per
$2$~\kms\ pixel at 3200\,\AA\ and $>$70\,pix$^{-1}$ between 4100 and
9200\,\AA.

It is desirable to combine all of the available exposures to maximise
the S/N. However, if the transitions we measure are unresolved by the
lower resolution $R\sim45000$ spectra, then using a combination of the
higher and lower resolution spectra could bias the derived absorber
parameters. Therefore we checked that relevant transitions are
resolved by the lower resolution exposures by making two combined
spectra, one from each subset of exposures with a common slit
width. There was no change in the absorption profiles of interest
between these two spectra, which indicates that the profiles are
indeed resolved by the lower resolution. We thus use a single combined
spectrum from all the exposures for our analysis.

\subsection{Measurement of column densities}

\begin{figure*}
\includegraphics[width=0.93\linewidth]{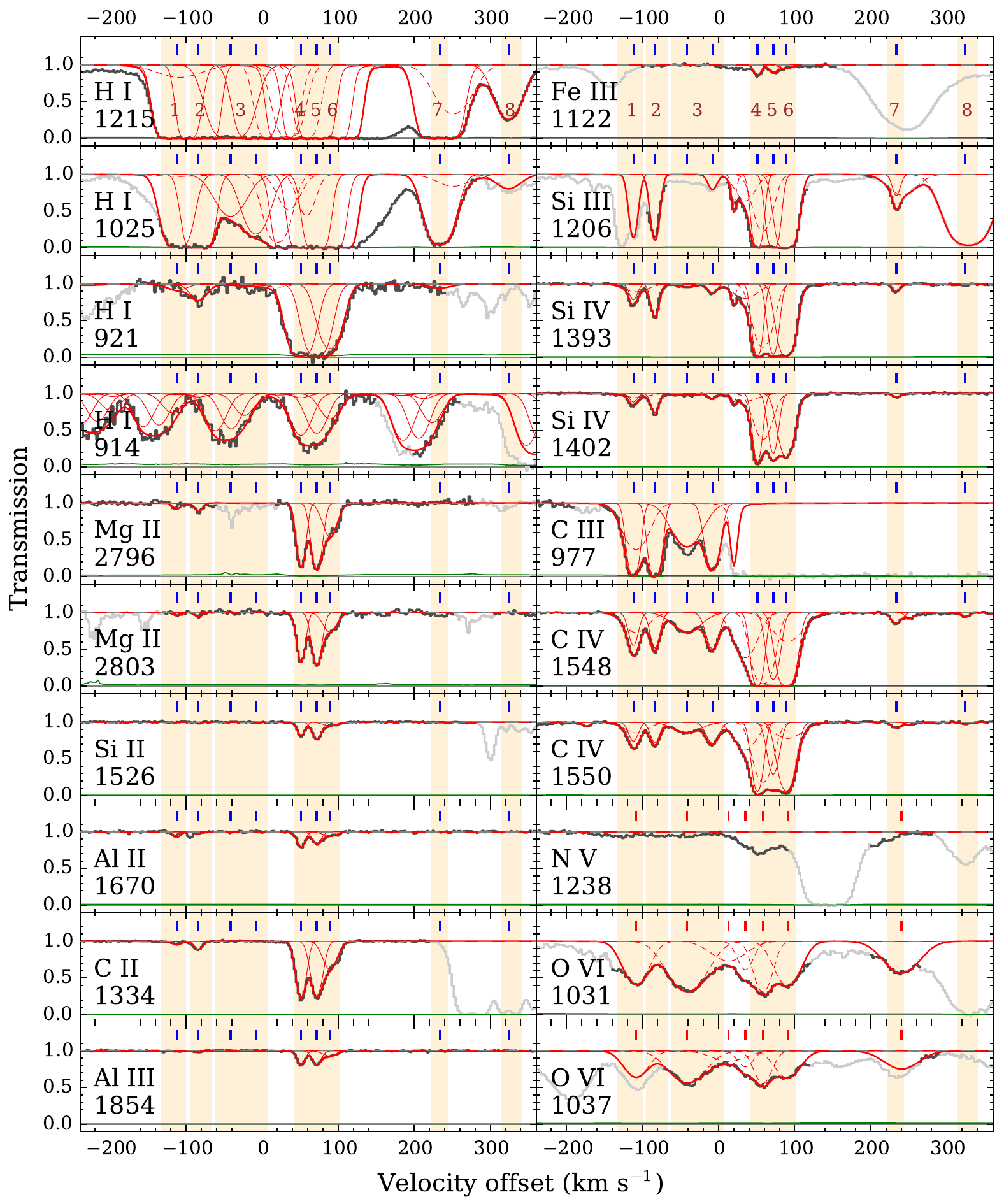}
\caption{\label{f_abs} Neutral hydrogen and metal transitions in the
  partial Lyman-limit system associated with the emitter. Zero
  velocity is at $z=2.4633$. This is $-220~\kms$ from the redshift of
  the galaxy \lya\ emission line, which is a typical \lya\ offset from
  the intrinsic redshift for LAEs. The thick red curve shows the model
  we fit to the UVES spectrum (dark histogram), and the thin green
  line near zero is the $1\sigma$ uncertainty in the
  transmission. Greyed out regions are blended with unrelated
  absorption. Thin red solid curves show the individual components we
  use in our \textsc{cloudy} modelling, and dotted curves are other
  components we do not model. The component numbers from Tables
  ~\ref{t_vpfit} and \ref{t_par} are shown in the top panels (note
  that component 3 is made up of two sub-components, shown by the 3rd
  and 4th ticks from the left). Metal transitions are seen over a
  velocity width of 436~\kms, and a single velocity structure, shown
  by blue ticks, can explain \HI\ and all the low-ion metal
  transitions. Different components are needed to fit \OVI; these are
  shown by red ticks in the bottom right panels.  We applied a
  $-1.2$~\kms\ offset ($0.6$ pixels) to the \SiIII\ absorbing
  region. The origin of this shift is unclear, but it may be caused by
  wavelength calibration uncertainties. \NV\ $\lambda$1242 is blended
  with forest absorption, and so we use \NV\ $\lambda$1238 to give an
  upper limit on $N_\mathrm{NV}$.}
\end{figure*}

The QSO spectrum reveals a partial Lyman-limit system
($\NHI=10^{16.94\pm0.10}$~\cmm) at $z=2.4639$, $\sim 150~\kms$
bluewards of the galaxy's \lya\ emission redshift.  \CII, \CIII, \CIV,
\SiII, \SiIII, \SiIV, \MgII, \FeIII, \AlII, \AlIII\ and
\OVI\ transitions are present, and there are eight clearly separable
absorption components that cover a velocity range of 436~\kms\ (see
Fig.~\ref{f_abs}). A single velocity structure can adequately fit all
the low-ions (which we define as having an ionization potential
$<3$~Ryd)\footnote{The ionization potential is the energy required to
  remove the outer electron of a species.}. \OVI\ roughly follows the
velocity structure of the lower ions, but its components are not
always precisely aligned with the lower ionization potential ions, and
the widths are much larger than would be expected if they were
produced by the same gas as the low-ions. \SiIV\ and \CIV\ show a
mixture of both broad components aligned with \OVI, and narrower
components aligned with the low ions.  We discuss the origin of the
\OVI\ absorption in Section~\ref{s_OVI}.

\begin{figure}
\includegraphics[width=\linewidth]{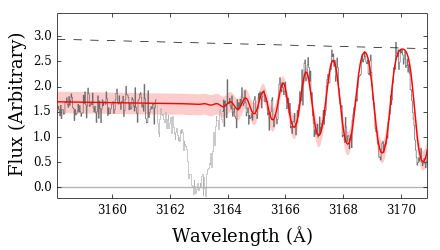}
\caption{\label{f_ll} The Lyman limit of the $z=2.466$ absorber. The
  solid curve and range shows a single component model with
  $\NHI=10^{16.94\pm0.10}~\cmm$, where the uncertainty includes a
  5\%\ error in the continuum level. The dashed line shows our
  adopted continuum level and the greyed out region is unrelated
  absorption.}
\end{figure}

Due to the high S/N and coverage of the Lyman limit (Fig.~\ref{f_ll}),
\NHI\ can be precisely measured in each low-ion component, enabling us
to place tight constraints on the metallicity and other physical
conditions of the gas using photoionization modelling. We measured
column densities by fitting Voigt profiles with
VPFIT\footnote{\url{http://www.ast.cam.ac.uk/~rfc/vpfit.html}};
parameters are given for each component in Table.~\ref{t_vpfit}.

\begin{table*}
\renewcommand{\arraystretch}{0.95} 
\addtolength{\tabcolsep}{-0.5pt}
\begin{center}
\begin{tabular}{ccccccccc}
\hline
 Ion & $z$ &  $b~(\kms)$ & $\log_{10} N~(\cmm)$ & \hspace{1cm} & Ion & $z$ &  $b~(\kms)$ & $\log_{10} N~(\cmm)$  \\
\hline
\multicolumn{5}{l}{{\bf Component 1} (velocity $=-112$~\kms)}             &      \multicolumn{4}{l}{{\bf Component 5} (velocity $= 72$~\kms)}     \\
 \HI    & 2.462033(02)& $17.4\pm 0.1$ &$ 14.88\pm 0.01$ & & \HI             &  2.464158(01)&  17.0$^d$      &$16.5 \pm     0.1   $  \\  
 \OI    &             &               &$ < 12.41  $     & & \OI             &              &                &$ < 12.77           $  \\
 \MgII  &             & 5.6$^a$       &$ 11.4\pm 0.2  $ & & \MgII           &              &  7.0$^b$       &$12.91 \pm   0.01 $    \\
 \SiII  &             &               &$ < 11.77  $     & & \SiII           &              &  6.75$^b$      &$12.85 \pm   0.01 $  \\  
 \SiIII &             &               &$ < 12.8   $     & & \SiIII          &              &  6.8$^b$       &$>13.75             $  \\
 \SiIV  &             & $5.5\pm1.0$$^a$& $12.22 \pm0.03$& & \SiIV           &              &  6.8$^b$       &$13.43  \pm  0.01 $  \\  
 \CII   &             & 7.1$^a$       &$ 12.22\pm 0.14$ & & \CII            &              &  $8.6\pm0.2^b$&$13.72 \pm   0.01 $  \\   
 \CIII  &             &               &$ >13.0        $ & & \CIII           &              &  6.8$^b$       &$ -                 $  \\
 \CIV   &             & 7.1$^a$       &$ 13.06 \pm0.02$ & & \CIV            &              &                &$13.74 \pm   0.04 $  \\  
 \AlII  &             & 5.5$^a$       &$ 11.1 \pm 0.2 $ & & \AlII           &              &  8.6$^b$       &$11.528 \pm   0.036 $  \\
 \AlIII &             & 5.5$^a$       &$ 10.8\pm  0.4 $ & & \AlIII          &              &  6.6$^b$       &$12.025 \pm   0.017 $  \\
 \FeIII &             &               &$ < 13.0   $     & & \FeIII          &              &  6.0$^b$       &$12.97\pm0.05       $  \\
 \NII   &             &               &$ < 13.28  $     & & \NII            &              &                &$ < 13.18           $  \\
 \NV    &             &               &$ < 12.8   $     & & \NV             &              &                &$ < 13.6            $  \\
 \OVI   &             &               &$ < 14.0   $     & & \OVI            &              &                &$ < 14.0            $  \\
\multicolumn{5}{l}{{\bf Component 2} (velocity $= -84$~\kms)}             &         \multicolumn{4}{l}{{\bf Component 6} (velocity $=92$~\kms)} \\
 \HI    & 2.462361(01)&  $14.2\pm0.3$ &$15.232\pm0.012$  & &  \HI             &  2.464358(01)&  17.0$^d$         &$16.2 \pm       0.15$   \\
 \OI    &             &               &$ <12.4        $  &  & \OI             &              &                  &$ < 12.77          $   \\  
 \MgII  &             & 5.3$^a$         &$11.52 \pm 0.09$& &  \MgII         &              &  11.5$^c$        &$12.476  \pm  0.017$   \\    
 \SiII  &             &               &$ <11.83       $  & &  \SiII           &              &  11.5$^c$        &$12.343  \pm  0.016$   \\  
 \SiIII &             &               &$<12.9$           & &  \SiIII          &              &                  &$>13.5             $   \\  
 \SiIV  &             & 5.5$^a$       &$12.593\pm0.009$  & &  \SiIV           &              &   $11.5\pm0.1^c$&$13.675 \pm   0.005$   \\   
 \CII   &             & $6.4\pm0.9$$^a$&$12.56 \pm 0.04$ & &  \CII           &              &  11.5$^c$        &$13.37   \pm  0.01$   \\    
 \CIII  &             &               &$>13.2         $  & &  \CIII           &              &                  &$ -                $   \\  
 \CIV   &             & 6.4$^a$       &$13.097\pm0.009$  & &  \CIV            &              &   11.5$^c$       &$14.189 \pm   0.015$   \\  
 \AlII  &             & 5.2$^a$       &$10.64 \pm0.25 $  & &  \AlII           &              &   11.5$^c$       &$11.27  \pm   0.08 $   \\  
 \AlIII &             & 5.2$^a$        &$10.94 \pm0.18 $ & &  \AlIII         &              &   11.5$^c$       &$11.794 \pm   0.034$   \\   
 \FeIII &             &               &$ < 12.5       $  & &  \FeIII          &              &   11.5$^c$       &$12.84 \pm 0.09      $  \\ 
 \NII   &             &               &$ <12.2        $  & &  \NII            &              &                  &$ < 13.18          $   \\  
 \NV    &             &               &$ <12.8        $  & &  \NV             &              &                  &$ < 13.6           $   \\  
 \OVI   &             &               &$ <14.0        $  & &  \OVI            &              &                  &$ < 14.0           $   \\  
\multicolumn{5}{l}{{\bf Component 3} (velocity $=-45$ to $-5~\kms$)}      &         \multicolumn{4}{l}{{\bf Component 7} (velocity $=235$~\kms)} \\
 \HI    &2.462848(04),&               &$ 14.6  \pm    0.3  $  &  &  \HI       & 2.466027(04) &      $19.1\pm0.4$ &$ 14.70 \pm   0.01 $   \\ 
 \OI    & 2.463234(02)$^e$&               &$ < 12.59           $  &  &  \OI       &              &                   &$ <13.18           $   \\ 
 \MgII  &             &               &$ <11.71            $  &  &  \MgII     &              &                   &$ <11.40           $   \\ 
 \SiII  &             &               &$ < 11.97           $  &  &  \SiII     &              &                   &$ <11.39           $   \\ 
 \SiIV  &             &               &$ 12.28_{-0.11}^{+0.25}$  & &  \SiIII    &              &                   &$ <12.1            $   \\  
 \CII   &             &               &$ <12.06            $  &  & \SiIV     &              &      $5.7\pm0.9$  &$ 11.88 \pm   0.04 $   \\  
 \CIII  &             &               &$ 13.65 \pm  0.3    $  &  &  \CII      &              &                   &$ <12.36           $  \\  
 \CIV   &             &               &$ 13.52_{-0.08}^{+0.15}$  & & \CIV      &              &      $8\pm1$      &$ 12.53 \pm   0.04 $   \\  
 \AlII  &             &               &$ <11.02            $  &  &  \AlII     &              &                   &$ <10.84           $   \\ 
 \AlIII &             &               &$ <11.19            $  &  &  \AlIII    &              &                   &$ <11.11           $   \\ 
 \FeIII &             &               &$ < 12.7            $  &  &  \FeIII    &              &                   &$<14.50            $  \\  
 \NII   &             &               &$ < 12.47           $  &  &  \NII      &              &                   &$ <12.65           $   \\ 
 \NV    &             &               &$ <12.75            $  &  &  \NV       &              &                   &$ < 13.1           $   \\ 
 \OVI   &             &               &$ < 14.0            $  &  &  \OVI      &              &                   &$ <14.0            $   \\ 
\multicolumn{5}{l}{{\bf Component 4} (velocity $=51$~\kms)}               &         \multicolumn{4}{l}{{\bf Component 8} (velocity $=324$~\kms)} \\
 \HI    & 2.463919(01)&  $17.00$$^d$   &$16.5   \pm  0.1  $ & & \HI        &  2.467073(13)&  $21.1\pm0.3$     &$13.59 \pm   0.01  $  \\ 
 \OI    &             &                &$ < 12.77         $ & & \OI        &              &                   &$ <12.42            $  \\
 \MgII  &             &  5.7$^a$       &$12.824 \pm  0.013$ & & \MgII      &              &                   &$ <11.60            $  \\
 \SiII  &             &  5.5$^a$       &$12.68  \pm  0.02 $ & & \SiII      &              &                   &$ <11.74            $  \\
 \SiIII &             &                &$>13.5            $ & & \SiIII     &              &                   &$   -               $  \\
 \SiIV  &             &  5.5$^a$       &$13.630 \pm 0.016 $ &  & \SiIV     &              &                   &$ <11.19            $  \\
 \CII   &             &  $7.0\pm0.1$$^a$&$13.69 \pm   0.01 $& & \CII       &              &                   &$  -                $  \\
 \CIV   &             &  7.0$^a$       &$14.076 \pm 0.031 $ & & \CIV       &              &  $6.7\pm2$        &$12.01 \pm   0.08 $  \\  
 \AlII  &             &  5.5$^a$       &$11.59 \pm  0.03  $ & & \AlII      &              &                   &$ <10.82            $  \\
 \AlIII &             &  5.5$^a$       &$12.00 \pm  0.02  $ & & \AlIII     &              &                   &$ <11.10            $  \\
 \FeIII &             &  4.8$^a$       &$13.08\pm0.04     $ & & \FeIII     &              &                   &$<13.7        $  \\      
 \NII   &             &                &$ < 13.18         $ & & \NII       &              &                   &$  -                $  \\
 \NV    &             &                &$ < 13.6          $ & & \NV        &              &                   &$ < 13.7            $  \\
 \OVI   &             &                &$ < 14.0          $ & & \OVI       &              &                   &$ <14.0             $    
\end{tabular}
\begin{flushleft}
$^a$ $b$ value is tied for all metal species assuming turbulent broadening $b_{\rm turb}=4$~\kms.\\
$^b$ $b$ value is tied for all metal species assuming turbulent broadening $b_{\rm turb}=5$~\kms.\\
$^c$ $b$ value is tied for all metal species assuming purely turbulent broadening. \\
$^d$ $b$ value fixed. \\
$^e$ This component is made up of two sub-components with the listed redshifts. Column densities are for the sum of both sub-components.
\end{flushleft}
\vspace{-0.3cm}
\caption{\label{t_vpfit} The redshift, $b$ parameter, column density
  and their $1\sigma$ uncertainties for the 8 components shown in
  Fig.~\ref{f_abs}. Upper limits are calculated using the $5\sigma$
  equivalent width detection limit for undetected transitions, and by
  measuring the highest column density Voigt profile consistent with
  the data for blended transitions.}
\end{center}
\end{table*}

\subsection{Photoionization modelling}

Following previous analyses
\citep[e.g.][]{DOdorico01_LLS,Fox05,Simcoe06,Prochaska09_QPQIII,Fumagalli11_sci,Crighton13_cma},
we use the observed column densities to infer a metallicity and volume
density for each component by comparison to \textsc{cloudy}
photoionization model predictions. However, in contrast to these
previous analyses we add another free parameter to represent
uncertainty in the shape of the UV ionizing background. It has been
shown that in some cases, changing the shape of the ionizing spectrum
results in an order magnitude difference in the inferred metallicity,
and 0.5 dex in the inferred ionization parameter
\citep[e.g.][]{Fechner11}.  Previous analyses have approached this
problem by considering different discrete incident continua in
addition to the integrated UV background, such as an AGN spectrum
\citep[e.g.][]{Finn14}, or the ionizing spectrum expected from a
nearby galaxy \citep{Fox05,Simcoe06}. While this technique gives a
rough indication of the uncertainties introduced by assuming different
UV background shapes, it makes it difficult to estimate a
statistically robust range of metallicities consistent with the
observed column densities. This problem is compounded by the fact that
the energies important for ionizing the metal species we observe are
close to the \HI\ and \HeII\ Lyman limits, making it very difficult to
measure the shape of the ionizing spectrum for both QSOs (Lusso et
al., in preparation) and galaxies, the dominant contributors to the UV
background. Therefore all the UV background shapes used are uncertain.

Our approach to this problem is to introduce a parameter
$\alpha_\mathrm{UV}$ that changes the power-law slope of the incident
radiation field over the energy range $1-10$~Ryd, which is the most
important range for predicting column densities of the transitions we
observe.  Despite all the uncertainties mentioned, a Haardt-Madau
integrated background \citep[][hereafter HM12]{Haardt12} is
often found to broadly reproduce the column densities seen in low-ion
species in the circumgalactic medium
\citep[e.g.][]{Fox05,Stocke13,Werk14}. Therefore we take this as a
starting point, and introduce a parameter to add a tilt to this
fiducial spectrum between $1$ and $10$~Ryd. More details are given in
Appendix~\ref{a_Cloudy}.

\textsc{cloudy} version 13.03\footnote{\url{http://www.nublado.org}}
\citep{Ferland13} was used to generate a grid of photoionization
models as a function of the hydrogen volume density \nH, metallicity
$Z$, neutral hydrogen column density \NHI\ and a parameter
$\alpha_\mathrm{UV}$ which describes the tilt in the ionizing
spectrum. The \textsc{cloudy} models assume that the absorbing gas is
a thick slab illuminated on one side by a radiation field. The effects
of cloud geometry are only relevant when self-shielding becomes
important. We will show that this system is highly ionized, so
\HI\ self-shielding should not be significant. Self-shielding by
\HeII\ for photons $>4$~Ryd still could be substantial, so geometrical
effects could act to cause small deviations between the observed and
predicted column densities for high-ions. We assume there is no dust
and that elements are present in solar abundance ratios. A large
amount of dust in the absorber would result in a depleted Fe to Si
ratio relative to solar. The \FeIII\ absorption we measure is
consistent with our photoionization predictions (which assume solar
abundances), so we see no evidence for strong dust depletion.  We do
not use non-equilibrium ionization models from
\citet{Oppenheimer13_noneq}, which would be important if there were a
nearby AGN which has recently turned off. We have no evidence that
this is the case for this system, but cannot rule it out.  As
described in the appendix, we account for any minor deviations from
the equilibrium predictions caused by non-equilibrium effects,
relative abundance variations, and geometric effects by adopting a
minimum uncertainty in the measured column densities of 0.1 dex.

We construct a likelihood function using the observed column densities
and grid of \textsc{cloudy} models, including upper and lower limits. Limits
are treated as one-sided Gaussians with $\sigma$ given by a constant
value of $0.05$. Then we apply priors and generate posterior
distributions for the \nH, $Z$, \NHI\ and $\alpha_\mathrm{UV}$ using the
Markov Chain Monte Carlo (MCMC) code \textsc{emcee} \citep{ForemanMackey13},
using 4-D cubic interpolation on the grid of \textsc{cloudy} models.

\subsection{Photoionization modelling results}

\begin{figure*}
\includegraphics[width=0.7\linewidth]{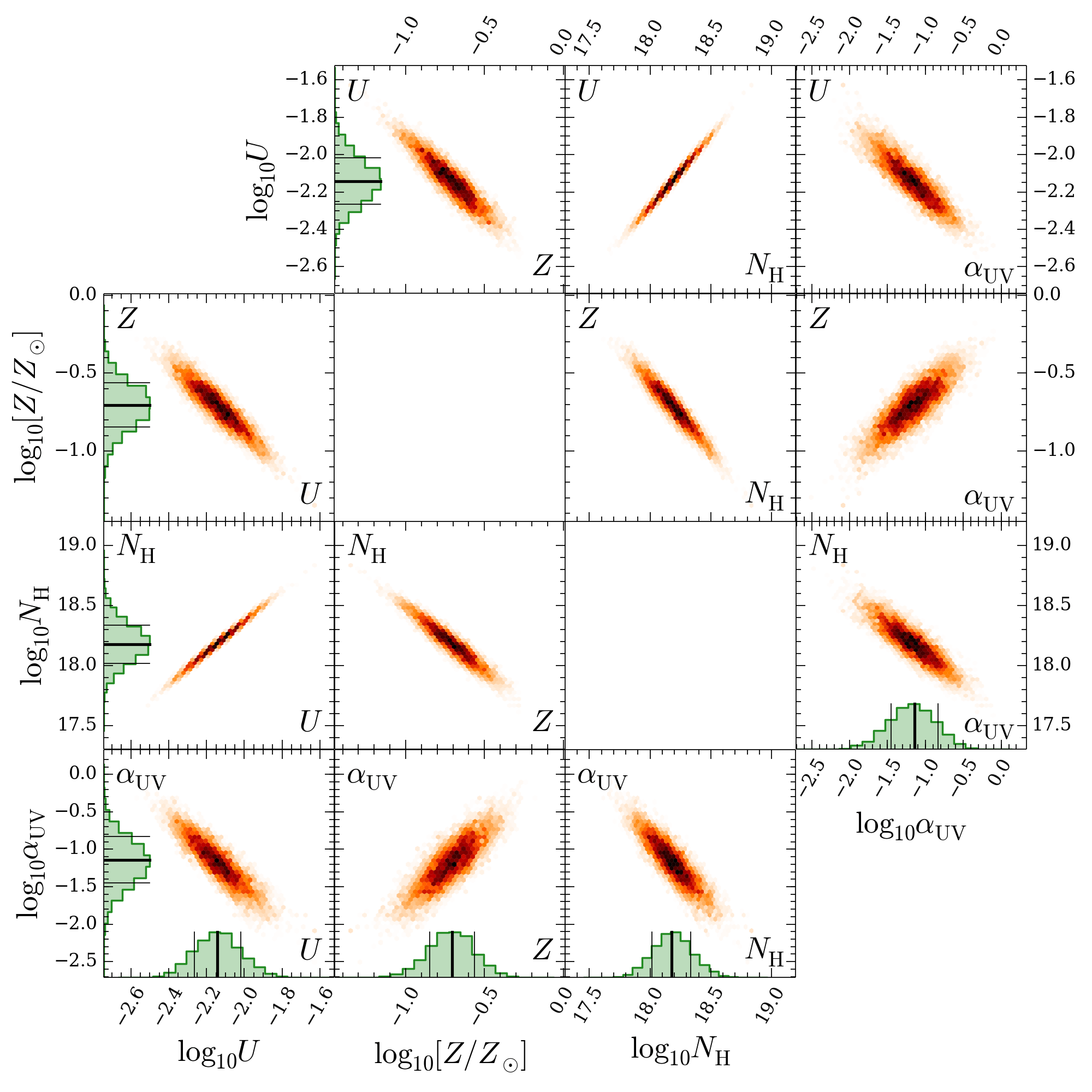}
\caption{\label{f_pos1} Posterior distributions for component 1. The
  distribution of MCMC samples is shown as a function of $\alpha_{\rm
    UV}$, \NH, $U$ (inversely proportional to \nH) and $Z$. Histograms
  show the marginalised sample distributions for each parameter, and
  the thin horizontal and vertical lines show the smallest interval
  containing 68\% of these samples.  Note that the top right panels
  are mirror images of the bottom left panels.}
\end{figure*}

\begin{figure}
\includegraphics[width=1.04\linewidth]{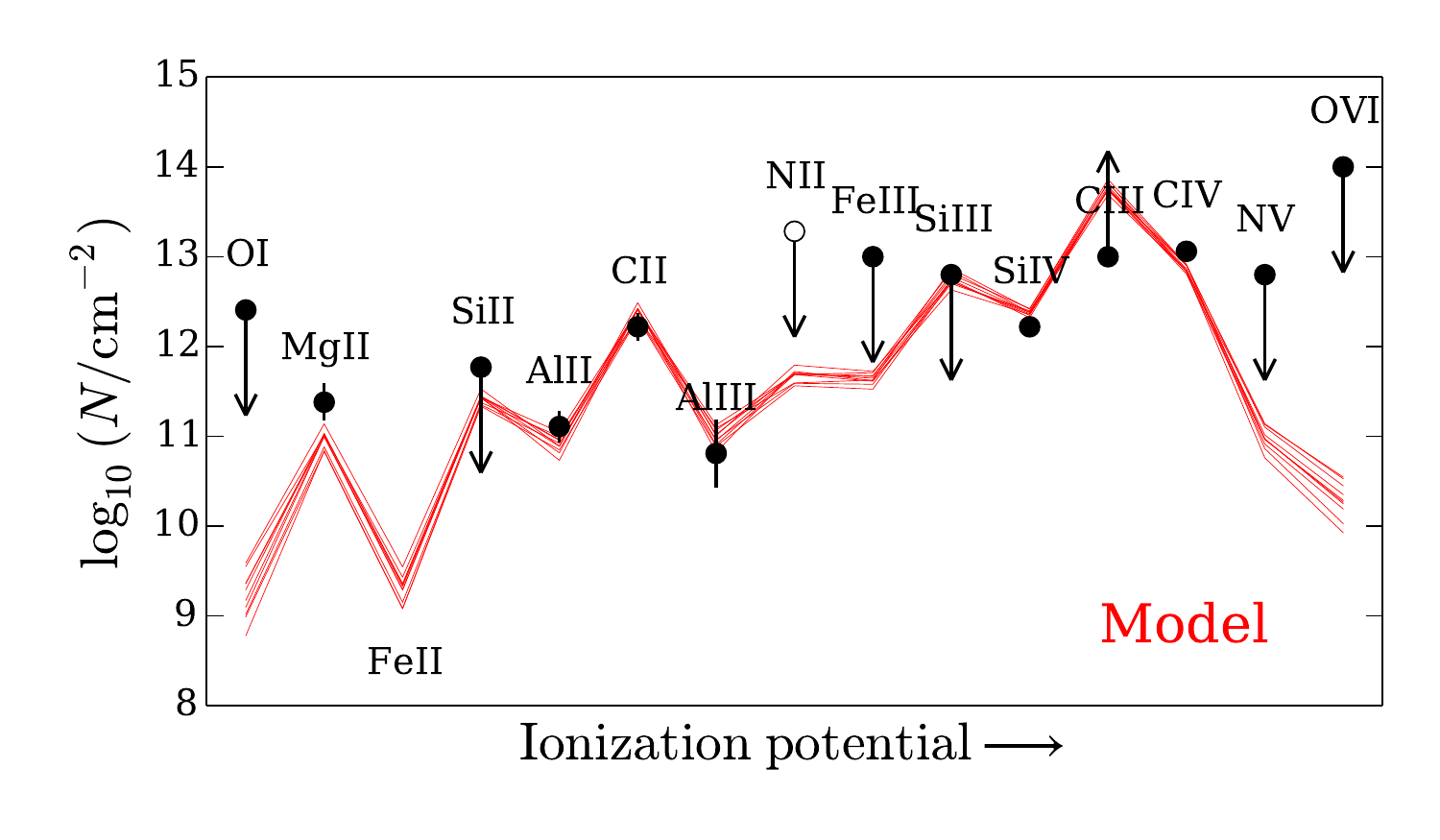}
\caption{\label{f_model1} The observed column densities for component
  1 and the predicted \textsc{cloudy} column densities for ten MCMC
  samples selected at random. Transitions are in order of increasing
  ionization potential, from left to right. Error bars show the $1
  \sigma$ uncertainties on the measured column densities. Open circles
  show observed values which were not used in the parameter estimation.}
\end{figure}

\begin{figure}
\includegraphics[width=1.04\linewidth]{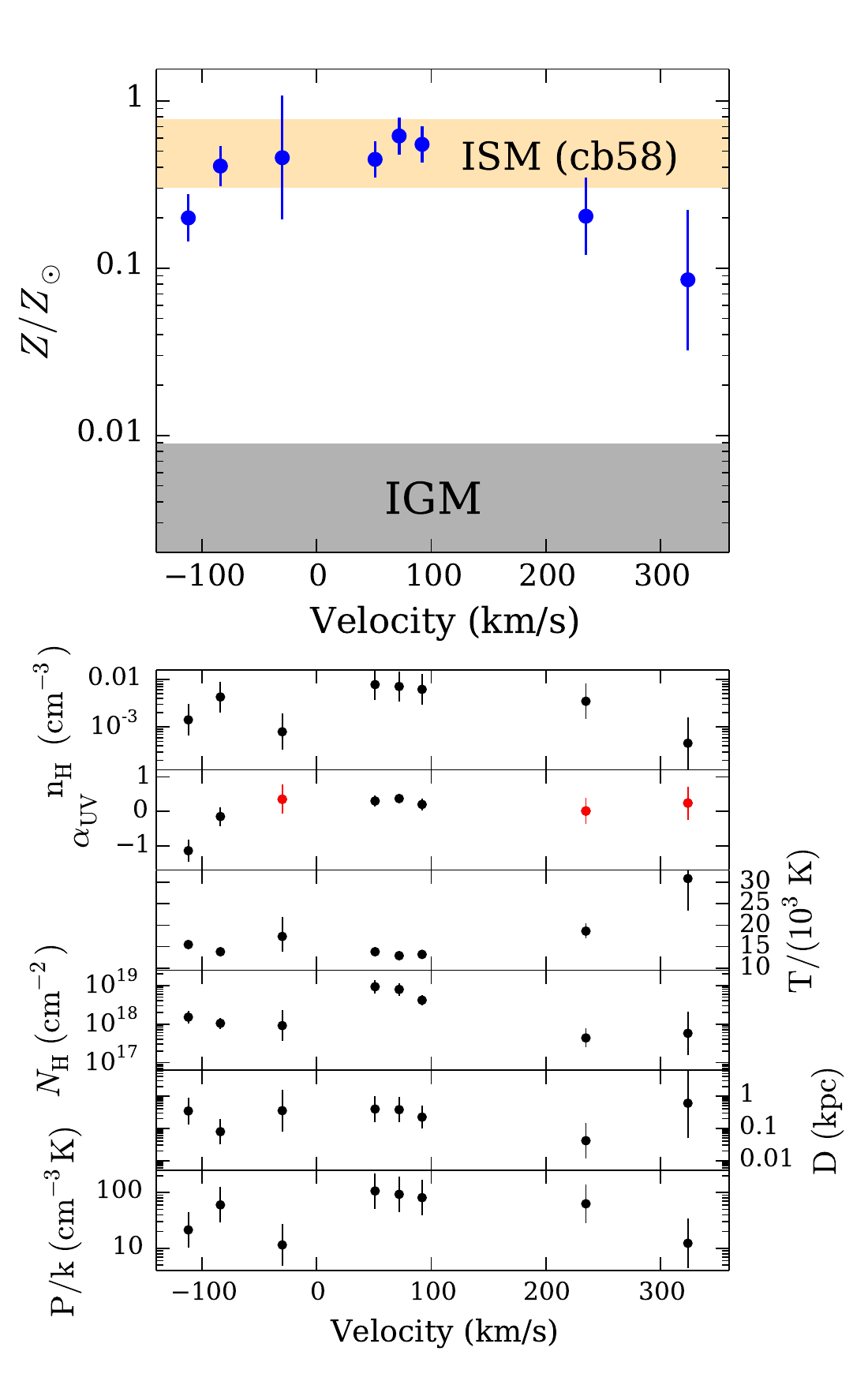}
\caption{\label{f_par} Parameters in Table~\ref{t_par} plotted as a
  function of component velocity. The top panel shows the metallicity
  in comparison to the measurement of the outflowing ISM metallicity
  in the lensed Lyman break galaxy CB58 \citep{Pettini02_cb58} and the
  metallicity of the background IGM at $z=2.5$ \citep{Simcoe04,
    Schaye03}. The lighter, red points in the $\alpha_\mathrm{UV}$
  plot show components where a prior of $\alpha_\mathrm{UV} = -0.25
  \pm 0.5$ has been applied.}
\end{figure}

Example outputs for the parameter estimation for a single component
are shown in Fig.~\ref{f_pos1} \& \ref{f_model1}. Fig.~\ref{f_pos1}
shows the posterior distributions, and Fig.~\ref{f_model1} shows the
observed column densities, and the predictions for ten MCMC parameter
samples selected at random.  Similar figures for all components are
shown in Appendix~\ref{a_Cloudy}. The inferred parameters and their
uncertainties are shown in Table~\ref{t_par} and Fig.~\ref{f_par}. The
cloud size is estimated as $\NH/\nH$, and the mass as
$4\pi/3(3D/4)^3\nH\mu m_p$, where $\mu m_p$ is the mass per hydrogen
atom, and we assume spherical clouds of radius $3D/4$.

We apply a Gaussian prior in log space to \NHI\ corresponding to the
measured column density. For components 1, 2, 4, 5 and 6 the
metallicity, spectral tilt and volume density can be constrained
independently. For the weaker components 3, 7 and 8, fewer metal lines
are detected and these three parameters cannot be independently
constrained.  Therefore we impose a prior distribution on
$\alpha_\mathrm{UV}$ of a Gaussian centred at $-0.25$ with
$\sigma=0.5$. This prior is consistent with the $\alpha_\mathrm{UV}$
range measured in the other 5 components.


The models can reproduce column densities of all the species apart
from \OVI, including the majority of \CIV\ and \SiIV. This suggests
that all of this absorption is produced by the same phase, and that
there is not a large fraction of \CIV\ associated with the gas
producing \OVI. The largest discrepancy between the models and data is
for \MgII, which is low by $0.2$--$0.4$ dex in the models for
components 1, 4, 5 and 6 (see Fig.~\ref{f_model1} and
\ref{f_model4}--\ref{f_model6}). We also observed a \MgII\ enhancement
in \citet{Crighton13_cma}, hinting that this could be a common feature
of the $z \sim 2.5$ CGM. This is unlikely to be caused by more
complicated variations in the UV background than those captured by
$\alpha_\mathrm{UV}$, as \SiII, which has a similar ionization
potential, is well-modelled in components 4, 5 and 6 (see
Figures~\ref{f_model4}, \ref{f_model5}, \ref{f_model6}). It could be
produced by a departure from solar abundance ratios. However, our
favoured explanation is that it is caused by a density variation in
the cloud (the \textsc{cloudy} models assume a constant density). If
there is a denser core near the centre of the cloud, this could result
in stronger very low-ion absorption relative to higher ions. Such
dense cores are known to exist in some high column density absorbers
at lower redshift \citep[e.g.][]{Crighton13_H2}. We intend to
investigate this scenario further in future work.

The gas metallicity ranges from $\log_{10}(Z/Z_\odot)=-1.1\pm 0.4$ for
component 8 to $\log_{10}(Z/Z_\odot)=-0.2\pm0.1$ for component 5. The
$\log_{10}U$ values are between $-2$ and $-3$, corresponding to
densities of $10^{-3}$ to $10^{-2}$~\cmmm\ (assuming $\Gamma_{\rm
  HI}=0.8\times 10^{-12}$~s$^{-1}$). These densities are similar to
the high-metallicity components in the absorber analysed by
\citet{Crighton13_cma}. As we describe in Appendix \ref{a_norm}, these
densities are lower limits. If there are any local ionizing sources in
addition to a Haardt-Madau UV background, these tend to result in
higher densities and thus smaller cloud sizes.  The component
metallicities are comparable to those measured in the outflowing
interstellar medium (ISM) of the lensed Lyman break galaxy
MS~1512-cb58 \citep{Pettini02_cb58}, and much larger than the IGM
metallicity at this redshift \citep{Schaye03, Simcoe04} as shown in
Fig.\ref{f_par}.

Interestingly, components 1, 4, 5 and 6 have a $\alpha_\mathrm{UV}$
different from zero, suggesting deviations from a HM12
background. Components 1 and 4/5/6 also have very different
$\alpha_\mathrm{UV}$ from each other, which may indicate an
inhomogeneous ionizing spectrum across the absorber. It is possible
that relative abundance variations or non-equilibrium effects could
mimic a non-zero $\alpha_\mathrm{UV}$. We have tried introducing
departures from solar abundance ratios, but find that the non-zero
$\alpha_\mathrm{UV}$ solutions remain. We defer a discussion on the
possibility of a variable ionizing radiation field and departures from
a HM12 background to a future paper. For this work, we simply treat
$\alpha_\mathrm{UV}$ as a nuisance parameter which is marginalised over
to find the metallicity and density.

If we use a fixed value of $\alpha_\mathrm{UV}$ we underestimate the
uncertainty on the metallicity by a factor of three. This demonstrates
the importance of including a variation in the UV slope when
estimating the metallicity.

\begin{table*}
\renewcommand{\arraystretch}{1}
\addtolength{\tabcolsep}{-1pt}
\begin{center}
\begin{tabular}{ccccccc}
\hline
      & vel. &     &                  & & $\log_{10} \NHI$ & $\log_{10} \NH$\\
comp.  &(\kms) & $\log_{10}(Z/Z_\odot)$& $\alpha_{UV}$ & $\log_{10} U$ & (\cmm)      &  (\cmm)   \\

\hline
 1 & $-112$ &$-0.70\pm0.14$ &$-1.14\pm0.31$ &$-2.14\pm0.12$ &$\mathbf{14.88\pm0.01}$ &$18.18\pm0.16$  \\
 2 & $-84$  &$-0.39\pm0.12$ &$-0.15\pm0.27$ &$-2.54\pm0.09$ &$\mathbf{15.23\pm0.01}$ &$18.02\pm0.14$  \\
 3 & $-30$  &$-0.34\pm0.37$ &$\mathbf{-0.35\pm0.42}$ &$-1.83\pm0.19$ &$\mathbf{14.47\pm0.26}$ &$17.96\pm0.40$  \\
 4 & +51   &$-0.35\pm0.11$ &$0.30\pm0.15$  &$-2.73\pm0.12$ &$\mathbf{16.43\pm0.09}$ &$18.97\pm0.17$  \\
 5 & +72   &$-0.21\pm0.11$ &$0.37\pm0.14$  &$-2.70\pm0.11$ &$\mathbf{16.37\pm0.09}$ &$18.90\pm0.16$  \\
 6 & +92   &$-0.26\pm0.11$ &$0.20\pm0.18$  &$-2.66\pm0.10$ &$\mathbf{16.02\pm0.10}$ &$18.62\pm0.13$  \\
 7 & +235  &$-0.69\pm0.23$ &$\mathbf{0.01\pm0.38}$  &$-2.43\pm0.17$ &$\mathbf{14.70\pm0.01}$ &$17.64\pm0.25$  \\
 8 & +324  &$-1.07\pm0.42$ &$\mathbf{-0.24\pm0.48}$ &$-1.63\pm0.44$ &$\mathbf{13.59\pm0.01}$ &$17.76\pm0.57$  \\[\smallskipamount]

\hline

 & vel. & $\log_{10} T$  & $\log_{10} \nH$ &$\log_{10} (P/k)$& $\log_{10} D$ & $\log_{10} M$ \\
comp. & (\kms)&  (K)      & (\cmmm)    & (\cmmm K)  &  (kpc)   & (\msun)  \\

\hline

 1 & $-112$ &$4.19\pm0.03$ &$-2.85\pm0.33(0.14)$&$1.33\pm0.32(0.12)$ &$-0.46\pm0.42(0.30)$ &$3.55\pm0.96(0.75)$  \\
 2 & $-84$  &$4.14\pm0.02$ &$-2.37\pm0.32(0.12)$&$1.78\pm0.32(0.10)$ &$-1.10\pm0.39(0.25)$ &$2.07\pm0.88(0.64)$  \\
 3 & $-30$  &$4.24\pm0.10$ &$-3.10\pm0.38(0.23)$&$1.06\pm0.37(0.22)$ &$-0.45\pm0.65(0.57)$ &$3.40\pm1.64(1.53)$  \\
 4 & +51   &$4.14\pm0.03$ &$-2.11\pm0.33(0.13)$&$2.03\pm0.32(0.12)$ &$-0.40\pm0.41(0.28)$ &$4.45\pm0.94(0.72)$  \\
 5 & +72   &$4.11\pm0.03$ &$-2.15\pm0.32(0.12)$&$1.97\pm0.32(0.12)$ &$-0.42\pm0.39(0.26)$ &$4.34\pm0.89(0.66)$  \\
 6 & +92   &$4.12\pm0.03$ &$-2.21\pm0.32(0.11)$&$1.91\pm0.32(0.12)$ &$-0.65\pm0.36(0.21)$ &$3.59\pm0.80(0.53)$  \\
 7 & +250  &$4.27\pm0.04$ &$-2.46\pm0.37(0.21)$&$1.80\pm0.35(0.17)$ &$-1.38\pm0.55(0.46)$ &$1.17\pm1.31(1.16)$  \\
 8 & +324  &$4.49\pm0.12$ &$-3.34\pm0.55(0.46)$&$1.09\pm0.45(0.34)$ &$-0.22\pm1.08(1.04)$ &$3.60\pm2.71(2.64)$  \\[\smallskipamount]

\end{tabular}
\caption{\label{t_par} Parameters estimated directly by MCMC sampling
  ($Z$, \nH, $\alpha_{\rm UV}$ and $\NHI$) and other derived
  parameters for the 8 components we model. The metallicity has a flat
  prior between $-3$ and $0.5$, the number density has a flat prior
  between $-4$ and $0$, and $\alpha_\mathrm{UV}$ has a flat prior
  between $-3$ and $2$. Bold indicates that value has an additional
  prior applied: for $\alpha_\mathrm{UV}$ this is a Gaussian
  distribution centred on $\alpha_\mathrm{UV}=-0.25$ with
  $\sigma=0.5$, and for \NHI\ it is the \NHI\ measurement in that
  component.  The uncertainties are $1\sigma$, marginalised over all
  other parameters, and so take into account covariances among the
  parameters. We include an additional 0.3 dex systematic uncertainty
  in $\nH$ due to the unknown normalisation of the incident radiation
  field. This is propagated to all quantities which are derived from
  $\nH$: $D$, mass and $P/k$. For these quantities, the uncertainty
  without including this $\nH$ systematic is shown in parentheses. }
\end{center}
\end{table*}

\section{Physical conditions in the galaxy's CGM}
\label{s_models}

We have shown that the partial LLS at an impact parameter of 50 kpc
from the galaxy is metal enriched to a level seen in the ISM of Lyman
break galaxies, far above the IGM metallicity. The galaxy is faint
(0.2 L$^*$) and likely has a low halo mass ($\sim10^{11.4}\msun$). All
other $R < 25.5$ BX candidates within $15\arcsec$ have been targeted
with spectroscopy and ruled out as being $<1000$~\kms\ from the LAE
redshift (Cooksey et al., in preparation, Crighton et al., in
preparation). Therefore there is no evidence that this is a satellite
in a higher-mass halo.  The QSO point spread function covers a small
solid angle, but it is possible it may hide another $z=2.5$ galaxy
even closer the QSO sightline ($<1.5\arcsec$, or $<10$~kpc). Higher
resolution imaging would help to identify any closer galaxy candidates
to the QSO sightline.

Even though this is a single galaxy-absorber pair, because this is the
first LAE studied in absorption and the pair was not absorption
selected, the covering fraction of gas around similar galaxies is
likely to be significant. The binomial $95\%$ confidence range for one
success from a single observation is 0.2--1, suggesting a covering
fraction $>20\%$\footnote{There could be a publication bias here:
  other groups may have discovered a similar galaxy close to a QSO
  sightline without an absorber and declined to publish the result. We
  are not aware of any unpublished systems, but this effect is
  difficult to quantify.}.  This is consistent with the association on
$\sim50$~kpc scales between \MgII\ and low stellar mass galaxies at
$z\approx1$--$2$ in \citet{Lundgren12}. A large covering fraction of
strong \HI\ is also seen in higher mass halos, around QSOs at $z\sim2$
\citep{Prochaska13_QPQVI} and around BX-selected galaxies at
$z\sim2.3$ \citep{Steidel10, Rudie12}. We note the rest equivalent
width of \MgII\ 2796 is 0.37~\AA, which lies close to the empirical
$\rho-W_r$ relation found by \citet{Chen10_emp} at $z<0.5$. This is
consistent with the idea that the circumgalatic medium extent remains
roughly constant from $z\sim 2.5$ to $z<0.5$ \citep{Chen12}.

At redshifts $<1$ \citet{Lehner13} have discovered a metallicity
bimodality for systems with a similar \NHI\ ($10^{16
}$--$10^{19}$~\cmm) to our partial LLS. The existence of a metallicity
bimodality at $z>1$ has not yet been established, but it is
interesting that the partial LLS component metallicities we measure
are all consistent with the high-metallicity branch of this
bimodality, which Lehner et al. interpret as being produced by
outflowing winds, stripped gas or recycled outflows. In the following
sections we explore possible physical origins for the partial LLS.

\subsection{Can the absorption be caused by IGM gas?}

What is the origin of the photoionized gas we measure? First we
consider whether these clouds have the physical properties of
\lya\ forest absorbers, where the Jeans scale in the photoionized IGM
determines their size and structure. \citet{Schaye01_lyaf} presented a
formalism for optically thin clouds---as is the case here---where the
thermal pressure of the cloud is balanced by its gravitational
pressure. This formalism relates the cloud's \NHI\ to \nH:
\begin{equation}
\begin{split}
\NHI = 2.3 \times 10^{13}\,\cmm \left(\frac{\nH}{10^{-5}\,\cmmm}\right)^{3/2} \left(\frac{{\it T}}{10^4\,\mathrm{K}}\right)^{-0.26} \\
\left(\frac{\Gamma}{10^{-12}\,\mathrm{s^{-1}}}\right)^{-1} \left(\frac{f_g}{0.16}\right)^{0.16}
\end{split}
\end{equation}
\noindent where $\Gamma$ is the hydrogen photoionization rate, $T$ is
the gas temperature, and $f_g$ is the gas mass fraction. Using the
\nH\ measurements for our absorbers predicts column densities that are
one to two orders of magnitude larger than the measured values. Our
volume density estimates scale linearly with the incident radiation
normalisation, so if our assumed UV incident radiation is
underestimated by 10--100 times this disagreement could be
eased. However, a factor of ten increase in UV radiation over the UVB
cannot be produced by star-formation in the nearby galaxy, which is
$\lesssim 10\msunyr$. We conclude that the gas does not arise in the
ambient IGM. This conclusion is reinforced by the gas metallicity,
which is much higher than expected for the IGM.

\subsection{Pressure confinement}

Are the clouds pressure confined by a hotter gas phase
\citep[e.g.][their section 5.2]{Prochaska09_QPQIII}? Some models of the
CGM predict that cool gas is condensed from a hot halo via
hydrodynamical instabilities
\citep[e.g.][]{Mo96,Maller04,Kauffman09}. In these models, the
\OVI\ absorption we observe is naturally explained as a warm envelope
around the cool clumps caused by their interactions with the hotter
halo.  In this case the cool gas should be in pressure balance with
the hotter halo, which is close to the virial temperature
($\sim7\times10^5$~K for a $10^{11.4}\msun$ halo).  The pressure of
each component of our absorber is well constrained, and $P/k$ ranges
from $10$ to $100$~\cmmm\,K. To match the pressure of the cool clouds,
this hotter phase would need to have a volume density:
\begin{equation}
n_\mathrm{H, hot} = 10^{-4.1}\,\cmmm \left(
\frac{{\it T}_{hot}}{7\times10^5\,K}\right)^{-1} \left(
\frac{{\it P/k}}{50\, \cmmm K}\right)
\end{equation}
Assuming this hotter phase fills the halo with constant density to
$r=50$~kpc, it would have a total mass $1.4\times10^{9}$\msun. This is
comparable to the galaxy's likely stellar mass,
$2\times10^{9}$\msun\ using the stellar-mass/halo-mass relation from
\citet{Moster13}. It is also a lower limit, as we expect the density
to follow the dark matter profile, and the gas halo is unlikely to be
truncated at 50~kpc. Therefore if the cool gas we see is pressure
confined, then the CGM must contain hot gas with total mass comparable
to the stellar mass of the galaxy. However, even if the gas is
pressure confined, the large velocity width (436~\kms) cannot be
readily explained by models of cold gas condensing from a static halo
via hydrodynamic instabilities.  In such models we would expect the
velocity width to be comparable to the circular velocity of a
virialized Navarro-Frenk-White halo, $100~\kms$ for
$M_\mathrm{halo}=10^{11}\msun$, or $\sim 220~\kms$ for
$M_\mathrm{halo}=10^{12}\msun$.

Therefore the gas we observe may be in pressure equilibrium with an
ambient hotter gas halo, but it is unlikely to have condensed out of a
static hot halo.

\subsection{Inflowing or tidally stripped gas model}

If the gas has not condensed from a hot halo, how did it find its way
into the CGM?  Could it have been pre-enriched at a higher redshift or
in a galactic fountain, and now be infalling towards the galaxy,
causing the velocity spread we see?  Or could it be stripped from the
ISM of fainter satellite galaxies?

We estimate a typical infall velocity along the QSO sightline of
\begin{equation}
v_\mathrm{infall} = \left(\frac{2 G M_\mathrm{halo}}{3R}\right)^{1/2}
\end{equation}
for a parcel of gas falling from infinity to a radius $R$ in a halo of
mass $M_\mathrm{halo}$. This can only explain the observed velocity
offsets if the halo mass is large ($10^{12}\msun$), which we have
argued is unlikely. In addition, the covering fraction of the small
absorbing gas clumps we see must be high. This follows from the small
individual cloud sizes (all $<1$~kpc), and the large relative
velocities of $200$--$400~\kms$ between clouds.  Given their small
size, each absorbing component cannot be physically associated, as
this would imply an enormous velocity shear. So individual components
are likely separated by distances similar to the impact parameter,
50~kpc, and their covering factor must be very high to intersect
several clumps in a single sightline (see \citealt{Prochaska09_QPQIII}
for a similar argument in relation to an absorber near a foreground
QSO).

Such a large covering fraction of small clumps is consistent with a
population of high velocity cloud (HVC)-like systems, similar to those
observed in the Milky Way's halo, some of which are infalling. On the
other hand, simulations generally predict that infalling gas is
manifest as a small number of narrow streams
\citep[e.g.][]{Fumagalli11,Shen13}, which seems incompatible with a
large covering fraction of small clumps.

Based on the high velocity width of the system we conclude that most
of the absorbing components are unlikely to be infalling or stripped,
but we cannot rule out some components being caused by inflowing gas.

\subsection{Outflowing gas model}

An alternative explanation is a supernovae-driven wind.  The distance
travelled by winds through the CGM is poorly constrained, but the
dependence of \MgII\ absorption with inclination angle around galaxies
at lower redshift suggests they could reach as far as $\sim 50$~kpc
\citep{Bordoloi11, Bouche12, Kacprzak12_azimuth}. In a wind model the
gas metallicity is naturally explained by recent supernovae
enrichment, and the large velocity width and high covering fraction is
due to an outflowing shell morphology. 

We can compare the absorber metallicity to that estimated using
emission from \HII\ regions in $z \sim 2.5$ galaxies from
\citet{Erb06_Z}. These metallicities are subject to greater systematic
uncertainties than those measured from absorption, because only a
small number of emission lines can be detected and the metallicity
indicators are calibrated at low redshift, but they provide the only
direct measurements of gas-phase metallicities in high redshift
galaxies. The mass-metallicity relation from \citet{Erb06_Z} indicates
the metallicity for a $z \sim 2$, $M^*=10^{9.1}$\msun\ galaxy is
$\sim0.3$ solar, similar to the absorber metallicities we measure.

We can also compare to metallicities which are directly measured in
outflowing winds from two $z=2.7$ lensed Lyman break galaxies. These
lensed galaxies are bright enough to be observed at moderate spectral
resolution, enabling the \HI\ and metal column densities to be
measured in the blueshifted ISM absorption. The first galaxy is cb58
at $z=2.7$, which has an outflowing gas metallicity of $\sim0.4$ solar
\citep{Pettini02}. The second is the ``8 o'clock arc'', another
$z=2.7$ lensed galaxy \citep{DessaugesZavadsky10} which has an
outflowing gas metallicity of 0.4--0.7 solar. These are both similar
to the metallicities we measure in the partial LLS. The 8 o'clock arc
outflow metallicity may be higher than that of cb58 and of several of
the partial LLS components, but the 8 o'clock arc is a highly luminous
galaxy with stellar mass $\sim4\times10^{11}$~\msun, much higher than
both cb58 ($\sim2\times10^{10}$\msun) and the galaxy in this
work. Therefore we expect its ISM metallicity to be $\sim 0.2$ dex
higher based on the mass-metallicity relation. We conclude that the
metallicities we find in the partial LLS are consistent with
metallicities directly measured in the outflowing winds of these two
$z=2.7$ Lyman break galaxies.

\begin{figure}
\includegraphics[width=\linewidth]{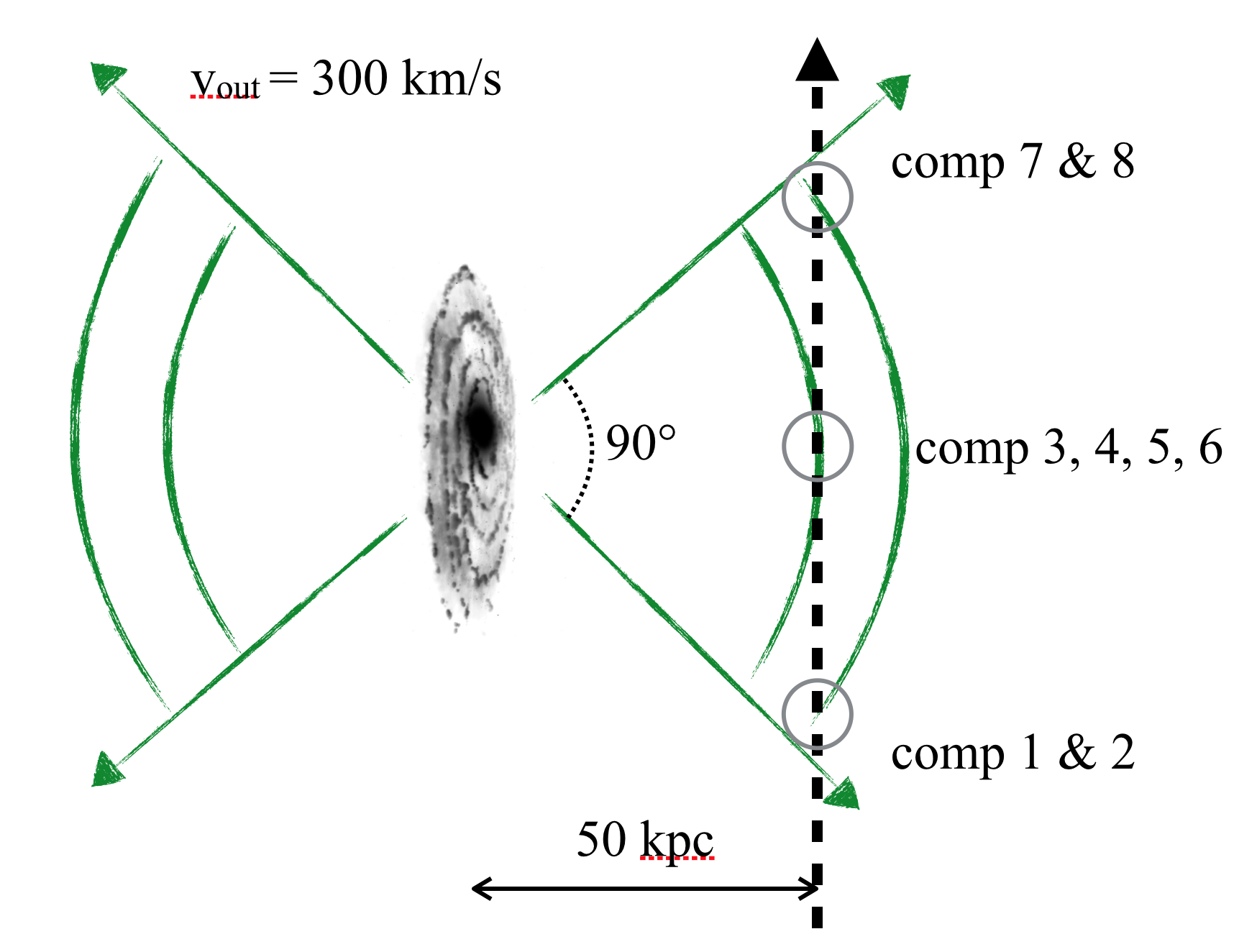}
\caption{\label{f_diag} A toy model of an outflow that is consistent
  with the absorption we observe. The dashed line is the sightline to
  the background QSO, which passes through a shell of gas outflowing
  from the nearby galaxy with velocity $300$~\kms. The density of the
  gas decreases with radius, meaning the strongest components are
  produced at the smallest radii from the galaxy. Here we assume an
  outflow opening angle of $90\deg$, but larger angles are also
  possible.}
\end{figure}

Fig.~\ref{f_diag} shows a possible outflow geometry where we have
intersected an outflowing cone whose density decreases with increasing
radial distance from the galaxy. Such a geometry is observed in lower
redshift outflows (e.g. M82, \citealt{Walter02}). Here a large
component of the wind velocity is perpendicular to the sightline. The
strong central components occur closest to the galaxy in a higher
density region of the outflowing shell, and the outer components are
produced in lower density regions more distant from the galaxy.

\subsubsection{Mass outflow rate}

In the outflow model we have a shell of outflowing gas with velocity
$v$ at radius $R$ which subtends a solid angle $\Omega$. Thus we can
estimate the mass outflow rate:
\begin{equation}
\begin{split}
\mathrm{\dot M} &= \mathrm{shell\ mass / travel\ time} \\
                &= \Omega\,  R^2  N_\mathrm{H,\perp} m_p \mu\,  v/R
\end{split}
\end{equation}
\noindent where $m_p\, \mu$ is the mass per hydrogen atom and
$N_\mathrm{H,\perp}$ the column density perpendicular to the shell
surface. We can generate the $200~\kms$ velocity offsets of the
outermost components (1, 2, 7 and 8) from the central components (4, 5
and 6) by assuming a geometry shown in Fig.~\ref{f_diag} and an
outflow velocity of $300~\kms$. This velocity is typical of winds
measured in brighter Lyman break galaxies at $z\sim2.3$
\citep{Steidel10}.  Using $N_\mathrm{H,\perp}=10^{19}~\cmm$,
$R=50$~kpc and $\mu=1.4$ gives
\begin{equation}
\label{e_mdot}
\begin{split}
\mathrm{\dot M}=5.4\msunyr\left( \frac{\Omega}{\pi}\right) \left(
\frac{v}{300\,\kms}\right)\\ \left( \frac{R}{50\, \mathrm{kpc}}\right)
\left( \frac{N_{H,\perp}}{10^{19}\,\cmm}\right).
\end{split}
\end{equation}
\noindent We have assumed a solid angle of $\pi$ steradians, which is
subtended by a biconical outflow with full-cone opening angles of
$80\deg$, as suggested by lower redshift studies using
\MgII\ \citep{Bordoloi11, Martin12}. Even though here we have just a
single sightline through the CGM, the large number of absorption
components suggests a large value of $\Omega$. A full-cone opening
angle of $140 \deg$, the largest permitted by low redshift studies
\citep{Rubin14}, would increase the outflow rate by a factor of
$3.5$. In this model a clumpy or filamentary outflow that is not a
smooth shell is accounted for by a reduced effective solid
angle\footnote{For a clumpy outflow, \NH\ in equation \ref{e_mdot} is
  the mean \NH\ over the shell area, averaging over
  clumps.}. Therefore any given outflow rate is consistent with a
larger opening angle if the outflow is clumpy.

Is the SFR of the galaxy consistent with driving such a wind?  The
galaxy is unresolved in the $r'$ imaging with $0.6\arcsec$ seeing,
implying it must have a size less than $\sim5$~kpc. Using the measured
SFR of $\sim1\msunyr$ and a half-light radius of 2.5~kpc (typical of
$z\sim2.5$ star-forming galaxies, e.g. \citealt{Law12}) gives
$\Sigma_\mathrm{SFR} = 0.05\msunyrkpcc$, comparable to the theoretical
value required to drive a superwind $\sim0.05\msunyrkpcc$
\citep{Murray11}. Lower redshift galaxies with similar star-formation
surface densities have also been observed to drive winds
\citep[e.g.][]{Rubin14}. The formalism presented by \citet{Murray05}
predicts the momentum deposition ($\mathrm{\dot{M}} v$) for a
radiatively-driven wind from a starburst. This formalism relates the
SFR to the momentum deposition by $\mathrm{\dot{M}} v = 2\times
10^{33}\ \mathrm{SFR}/(\msunyr)\,$g$\,$cm$\,$s$^{-2}$. Using our
measured mass outflow rate, the SFR required is $\sim5\,\msunyr$, five
times larger than the observed SFR. However, the observed SFR may not
have been the SFR in the galaxy at the time the wind was
launched. Starbursts have timescales of $0.1$--$1$~Gyr
\citep[e.g.][]{Thornley00,McQuinn10}. The time for a wind travelling
at 300~\kms\ to reach 50~kpc is $0.2$ Gyr, assuming the wind has not
deccelerated. Therefore it is possible that a previous starburst event
could have launched the wind, then subsequently died away leaving the
SFR that we measure. This scenario could be tested by taking a
rest-frame optical spectrum of the galaxy, which would show strong
post-starburst features, similar to a lower redshift galaxy that shows
a large-scale wind in absorption in a nearby QSO sightline
\citep{Tripp11}. Due to the faintness of the galaxy this is
challenging with current facilities, but will become straightforward
once the James Webb Space Telescope or 30m-class ground-based
telescopes become available.

This outflow rate is much smaller than the rate reported by
\citet{Steidel10} at $z\sim2.3$ around BX-selected galaxies
($\sim230\msunyr$) and is bracketed by the different estimates from
\citet{Martin12} ($\sim 23\msunyr$) and \citet{Rubin14}
($\sim1\msunyr$) for a $z \sim0.7$ galaxies with a similar SFR to the
galaxy in this work. We caution that in those studies gas absorption
features are both saturated and unresolved, \NHI\ cannot be measured
and the outflow geometry is not well constrained. This means the gas
metallicity, dust depletion, ionization corrections, and thus these
outflow rates are uncertain by an order of magnitude or more. The high
S/N and resolution of our QSO spectrum allows precise derivations of
the \NHI, metallicity, dust depletion and ionization corrections, so
in the context of an outflow model the largest uncertainties in the
outflow rate are due to the wind geometry. However, our analysis uses
a single galaxy-absorber pair and substantial uncertainties in the
geometry remain.

A larger sample of systems is necessary to confirm the outflow rate we
infer here is typical of $z\sim2.5$ galaxies.

\subsubsection{Escape velocity}

The outflow velocity required by our toy model to reproduce the
kinematics of the absorption is 300~\kms\ at 50~kpc. If the galaxy
halo mass is large, $\sim 10^{12}\msun$, this is below the halo escape
velocity $\sqrt{2GM_\mathrm{halo}/R} = 410~\kms$. However, if the halo
mass is $10^{11.4}\msun$, as suggested by the faint continuum and high
\lya\ equivalent width, the escape velocity is only $210~\kms$. In
this case the gas we see will escape the potential well of the galaxy
to enrich the IGM.

Several lines of evidence point towards low-halo mass galaxies being
responsible for IGM enrichment \citep[e.g.][]{Madau01,Booth12}. These
imply that galaxies in very low mass halos, M$_{\rm halo} \ll
10^{11}\msun$, must eject metals to enrich the IGM at high redshift to
the level that is observed at $z\sim3$. The galaxy in this work may be
the first indication of a low mass galaxy at $z\sim 2.5$ driving
metal-enriched gas into the IGM. Near-infrared observations of the
galaxy are thus highly desirable to determine whether it has a low
stellar mass, as expected from its faint UV magnitude.

\subsection{The origin of the \OVI\ absorption}

\label{s_OVI}

The \OVI\ absorption is shown in the bottom right panels of
Fig.~\ref{f_abs}, and its parameters are listed in
Table.~\ref{t_o6}. While the \OVI\ velocity components are not always
precisely aligned with the low-ions, they do roughly follow the
low-ion components, suggesting a physical connection. Therefore the
gas producing \OVI\ is either created by the cool clumps interacting
with their environment---for example, by warm gas ablated from the
cool clumps as they move through a hot halo, which one possible origin
for \OVI\ absorption seen around HVCs near the Milky Way
\citep[e.g.][]{Richter06}---or else produced by the same starburst
event that generated the clumps. This also suggests the \OVI\ gas must
have a similar size scale ($\sim$kpc) and metallicity (0.1--0.6 solar)
to the lower-ionization gas.

The \OVI\ absorption is broader than expected from the low-ion
absorption and $N_{\rm OVI}$ is not reproduced by our photoionization
models. Therefore it must be in a different phase, distinct from the
one we model. This phase could be photoionized, in which case it must
have a lower density than the cool clumps to avoid producing large
amounts of low-ion absorption. However, \citet{Lehner14} analyse a
sample of 15 Lyman-limit absorbers with associated \OVI\ similar to
this system (but without information about the presence of a nearby
galaxy), and report a correlation between $b_{\rm OVI}$ and $N_{\rm
  OVI}$. That is, they do not find many high \NOVI, narrow lines,
instead high \NOVI\ systems tend to have a large linewidths. They
argue this is caused by \OVI\ being produced by gas cooling
radiatively from a hot ($T\sim 10^{5-6}$~K) temperature, rather than
in photoionization equilibrium, as suggested by
\citet{Heckman02}. Several of the \OVI\ components in our system have
$b_{\rm OVI}=20-30$~\kms, corresponding to $T_{\rm max}=(4$--$9)
\times 10^5$~K assuming purely thermal broadening, which is consistent
with this scenario.

\begin{table}
\renewcommand{\arraystretch}{1} \addtolength{\tabcolsep}{-0.5pt}
\begin{center}
\begin{tabular}{cccc}
\hline
 vel. (\kms) & $z$ &  $b~(\kms)$ & $\log_{10} N~(\cmm)$  \\
\hline
 $-109$&   2.462072(05) &  $ 20.8  \pm 0.8$ & $ 13.96 \pm 0.01$\\
 $-42$ &   2.462848(09) &  $ 30.8  \pm 1.1$ & $ 14.24 \pm 0.02$\\
 +13   &   2.463479(80) &  $ 30$$^a$          & $ 13.67 \pm 0.04$\\
 +35   &   2.463734(27) &  $ 13.0  \pm 2.7$ & $ 13.5 \pm 0.2$\\
 +58   &   2.464001(11) &  $14.4   \pm 1.7$ & $ 13.94 \pm 0.07$\\
 +90   &   2.464377(13) &  $21.1   \pm 1.7$ & $ 13.99 \pm 0.04$\\
 +240  &   2.466107(07) &  $30$$^a$           & $ 13.93 \pm 0.01$\\
\hline
\end{tabular}
\begin{flushleft}
$^a$ $b$ value fixed.
\end{flushleft}
\caption{\label{t_o6} The velocity relative to the galaxy redshift,
  the component redshift, $b$ parameter, column density and their
  $1\sigma$ uncertainties for the \OVI\ components shown in
  Fig.~\ref{f_abs}. The uncertainties are taken directly from VPFIT,
  and do not include errors in the continuum placement.}
\end{center}
\end{table}

\citet{Simcoe06} consider two models to explain the broad \OVI\ in
another, similar absorber at impact parameter of 115 kpc from a
$z=2.3$ galaxy. The first is an outflowing shock front driven by a
supernovae wind, and the second is infalling, pre-enriched gas that is
shock heated during infall on to the halo. The infall scenario is
unlikely for our absorber, because of the large velocity extent of the
\OVI\ (360~\kms), and because of the close velocity association
between it and the low ions.

We conclude that the \OVI\ is likely caused by a warm, $\sim 10^5$~K
gas envelope around the cool photoionized clumps. This may be a
component of the outflowing gas, or the result of an interaction of
the outflowing gas with a hotter halo that could pressure-confine the
clumps. It may be radiatively cooling, and collisionally ionized
rather than photoionized.

\subsection{Mass in the CGM}

We can estimate the mass contribution of gas phases traced by low ions
and the \OVI\ to the CGM of this galaxy. These mass estimates are
independent of the model we assume for the gas (i.e. wind versus
inflowing or stripped); they depend only on the gas covering fraction,
$f_c$. Given the radial extent of the gas, $R_\mathrm{max}$, and using
\NH\ measured for the photoionized gas, we find a mass
\begin{equation}
\begin{split}
M_{\rm cool}& = \NH\, m_p\, \mu\, \pi\, R_\mathrm{max}^2\, f_c \\
 &= 4.4 \times 10^8 \msun \left(\frac{{\it R}_{\rm max}}{50\,{\rm kpc}} \right)^2 \left(\frac{\NH}{10^{19}\,\cmm}\right) \left(\frac{\it f_c}{0.5}\right)
\end{split}
\end{equation}
where $m_p\mu$ is the mean particle mass and we take $\mu = 1.4$. 

The \OVI\ is produced by a separate gas phase, and so contributes an
additional mass. The \NH\ associated with this phase can be estimated
$\NH = \NOVI / f_{\rm OVI} \times (Z_{\odot}/Z)$, where $f_{\rm OVI}$
is the fraction of oxygen in O$^{5+}$ and $Z$ is the gas
metallicity. \citet{Lehner14} use the models of
\citealt{Oppenheimer13_noneq} (see also \citealt{Gnat07},
\citealt{Vasiliev11}) to show that the maximum $f_{\rm OVI}$ is 0.2
for gas at temperatures $> 10^{4.5}$~K. Taking the metallicity to be
close to that of the photoionized clumps, 0.5~$Z_{\odot}$, this
implies $\NH=10^{19}~\cmm$ for the \OVI\ phase, similar to the low-ion
gas. This echoes results at lower redshift. \citet{Fox13} found the
mass contribution for the \OVI\ phase in a sample of LLSs at $z < 1$
is also comparable to the mass contribution from low-ions.

Therefore the combination of cool photoionized gas and gas associated
with \OVI\ contains a baryonic mass of
$\gtrsim0.8\times10^9\msun$. Using our fiducial stellar mass for the
galaxy of $1.4\times10^9\msun$, this represents $\sim60\%$ of the mass
in stars. The amount of mass in the CGM could be larger if it extends
beyond 50~kpc, or if either $f_{\rm OVI}$ or the \OVI\ gas metallicity
is lower than we have assumed.

\subsection{Implications for the origin of Lyman-limit systems}

This absorber falls just below the commonly used criterion for LLSs
($\NHI>17.2$), but is likely to share physical characteristics with
them. The covering fraction of $z\sim2.5$ LLS around QSOs has been
shown to be high \citep{Hennawi06_QPQI,Prochaska13_QPQV}, and they
strongly cluster around the QSOs \citep{Hennawi07_QPQII,
  Prochaska13_QPQVI}. \citet{Prochaska13_QPQVI} show that these
results suggest that the majority of all LLSs could originate within 1
Mpc of the high-mass halos ($\gtrsim 10^{12}\msun$) QSOs are believed
to populate. \citet{Steidel10} argue that the extended CGM of
Lyman-break selected galaxies may also contribute a dominant fraction
of LLSs, based on low-ionization metal absorption extending to an
impact parameter of 100 kpc. \citet{Rauch08} argue that faint LAEs,
with line fluxes of a few $10^{18}$~\ergscmm, may be responsible for
much of the LLS population. There will be some overlap between these
samples (some LAEs may also be Lyman break galaxies, and may be found
close to QSOs, for example). Is there also room for a population of
LLSs associated with $\sim 10^{11.4}$ halo mass galaxies?

\citet{Fumagalli13} consider two simple models that reproduce the LLS
incidence to $z=3$. In the first, LLSs inhabit the CGM of all galaxies
within a halo above a minimum halo mass corresponding to a circular
velocity of 200~\kms\ (below which the CGM gas can easily escape),
with the CGM halo extent fixed to the virial radius at $z=2.5$. The
second uses a lower minimum mass cutoff, motivated by most LLS being
caused by accreting gas. Due to its faint rest-frame UV magnitudes,
the galaxy in this work is likely in a $10^{11.4}\msun$ halo, which
implies the gaseous CGM of more numerous, fainter galaxies also
contribute to the LLS population. This is qualitatively in agreement
with recent cosmological numerical simulations
\citep{Erkal14,Rahmati14} which predict that most LLSs inhabit halos
with M$_{\rm halo}$ $<10^{11}\msun$. The LLS absorption cross section
suggested by Fig.~\ref{f_diag} is larger than in these cosmological
simulations.  The suite of single-galaxy simulations by
\citet{Fumagalli14} also predicts a lower covering fraction
$f_c\sim0.15$ than we have assumed ($f_c=0.5$). Some of this
difference could be due to this system being a partial LLS: weaker
LLSs likely have larger CGM covering fractions
\citep[e.g.][]{Rahmati14}.  However, perhaps more importantly these
simulations cannot yet resolve the hydrodynamics relevant for the
movement of these clouds through the CGM, as we discuss in
Section~\ref{f_sim}.

The incidence rate of LLSs at $z=2.5$ is $\sim 0.3$
\citep{Fumagalli13}.  The contribution of faint galaxies in the range
$L^*=0.15$ to $0.3$ (close to the $0.2 L^*$ luminosity of our galaxy)
to the incidence rate of LLSs, $l(X)$, is given by
\begin{equation}
\begin{split}
l(X) &= \frac{c}{H_0} n_\mathrm{com} A_\mathrm{eff} \\ &=
0.06\left(\frac{n_\mathrm{com}}{3.7\, \mathrm{comoving\ Mpc^{-3}}}
\right) \left(\frac{f_c}{0.5} \right) \left(\frac{R}{50\,
  \mathrm{kpc}} \right)^2
\end{split}
\end{equation}
where $A_\mathrm{eff} = f_\mathrm{c} \pi R^2$ is the effective
absorption cross section in proper coordinates, and $f_\mathrm{c}$ is
the covering fraction. $n_\mathrm{com}$ is the comoving number density
of faint galaxies, where we have used the luminosity function from
\citet{Reddy09}\footnote{Using the number density for \lya\ emitters
  \citep{Hayes10} with a similar \lya\ luminosity to our galaxy does
  not change our conclusions.}.  We conclude that a population of LLSs
around faint galaxies with properties similar to the one in this work
does not exceed the observed incidence rate, and still allows a large
fraction of LLSs to be produced in brighter galaxies, and/or in the
extended environments of higher mass halos as found by
\citet{Prochaska13_QPQVI}.

\section{Implications of sub-kiloparsec scale clouds}

\label{s_sim}

The characteristic sizes we derive for these clouds
(Table~\ref{t_par}) are very small: $100$--$500$~pc. These sizes are
robust to modelling uncertainties, and include uncertainties in the
ionizing field normalisation. Such small sizes have important
implications for the propagation of these clouds through the CGM, and
for numerical simulations of the CGM.

\subsection{Cloud disruption by hydrodynamic instabilities}

In the discussion of the outflow model we made extensive comparison
between the cool gas absorption we see at $50$~kpc to the cool gas
that is seen in low-ion outflows `down-the-barrel', in absorption
against individual and stacked galaxy spectra \citep{Steidel10}.  Our
favoured outflow picture sees these cool, metal-enriched clumps
travelling through hot, virialized gas already in the halo. Therefore,
we expect these clouds to be disrupted by Kelvin-Helmholtz
instabilities on a timescale
\begin{equation}
\tau_{\rm kh} \approx 4.1\,\mathrm{Myr} \left(\frac{R_{\rm cl}}{0.4~\mathrm{kpc}}\right) \left(\frac{v}{300~\kms} \right)^{-1} \left(\frac{n_{\rm cl}/n_{\rm halo}}{10}\right)^{1/2}
\end{equation}
 for a spherical clump of radius $R_\mathrm{cl}$ and density
 $n_\mathrm{cl}$, travelling at $v$~\kms\ through a halo with density
 $n_\mathrm{halo} $\citep[e.g.][their section 6.3]{Agertz07,Schaye07}. We
 have assumed a cloud-halo density contrast of 10, but the timescale
 only weakly depends on this value. This timescale is more than an
 order of magnitude shorter than the 0.2 Gyr required for the cloud to
 travel $>50$~kpc from the galaxy. If these clumps were produced in an
 outflow, how then can they have travelled more than 50 kpc from the
 galaxy without being destroyed? We consider three possible solutions.

First, there may not be any existing hot gas halo. If this is a low
mass galaxy ($M_\mathrm{halo}<10^{11}$\msun), then a shock may not yet
have formed and halo gas may not be at the virial temperature
\citep{Birnboim03}. However, it seems unlikely that the halo is
completely free of any gas, and the zoomed simulations of
\citet{Fumagalli14} suggest that some hot CGM gas is in place by
$z\sim2$, even in $M_\mathrm{halo} \sim 10^{11.2}$\msun\ halos. The
\OVI\ gas we see also suggests a hot halo is present. Therefore this
explanation seems unlikely.

A second possibility is that the cool gas does not actually traverse
the distance from the galaxy to its current location in the
CGM. Instead it condenses just behind a shock front driven by hot gas
in the outflow once the temperature of the front drops enough for
radiative cooling to become efficient, as outlined by
\citet[][their section 5.2]{Simcoe06}. This scenario can also explain the
\OVI\ absorption as cooling gas found behind the shock front. 

Finally, there may be a mechanism that makes the clumps resistant to
hydrodynamic instabilities.  Such mechanisms have been explored to
explain the apparent longevity of HVCs in the Milky Way halo, which
share many characteristics with the absorbers we analyse: they are
small, cool clouds moving through a hotter halo. Confinement by
magnetic fields is one possibility with some observational support
\citep[e.g.][]{McClureGriffiths10}. The effect of magnetic fields on
clouds remains unclear, however, and some simulations suggest they may
even hasten cloud disruption \citep{Stone07}, or strongly suppress the
relative velocity between the cloud and surrounding ambient medium
\citep{Kwak09}. The mechanism favoured by \citet{Putman12} for
extending the lifetimes of HVCs is the presence of a warm envelope
around the cool clouds (observed as \OVI\ absorption around the HVCs)
that significantly extends its lifetime.  It does this by minimizing
both the temperature and velocity gradients between the cloud and the
halo, making it resistant to instabilities. We also see \OVI\ in this
absorber that is likely caused by a warm gas envelope around the
clumps, so this is a plausible explanation.

To our knowledge, numerical simulations have not yet been performed to
test whether such a warm envelope can indeed extend the cloud lifetime
by the necessary amount.

\subsection{Small gas clump sizes are a generic feature of the CGM}

\label{f_size}

The absorbing clumps' small size makes them challenging to resolve in
simulations of the CGM. First, we emphasize that there is a large body
of evidence indicating that low-ion metal absorbers and partial LLSs
have sizes much smaller than a kiloparsec. Using lensed QSO sightlines
separated by 10s of parsecs, \citet{Rauch99} showed that a $z \sim 2$,
$\NHI=10^{16}~\cmm$ system with low-ion metal absorption must have a
size $<50$~pc. \citet{Petitjean00} found \MgII\ systems at $z = 1-2$
must have sizes $<1$~kpc due to partial covering of lensed images of a
background QSO. \citet{Simcoe06} found sizes 10$--$100~pc for
components in a partial LLS at an impact parameter of 115~kpc from
another $z\sim2.3$ galaxy using photoionization modelling. By
inferring the gas density from \CII$^*$ absorption,
\citet{Prochaska09_QPQIII} found similar sizes for gas clumps in a LLS
in the CGM of a $z\sim 2.5$ QSO. \citet{Schaye07} find small sizes for
$z\sim2.5$ metal-rich absorbers selected by their strong \CIV\ and
weak \HI, again using photoionization modelling. Such small sizes are
also found at low redshift: HVCs, with similar \NHI\ to LLS, have
sizes $<100$~pc \citep{BenBekhti09} inferred from \HI\ 21cm images,
and photoionization modelling of gas around $z\sim 0.1$ galaxies
\citep{Stocke13, Werk14}, of $z<1$ LLSs \citep{Lehner13} and of weak
\MgII\ systems \citep{Rigby02} also yields small sizes. Therefore
these small clump sizes are a generic feature of both the high and low
redshift CGM.

Moreover, for the partial LLS in this work it is not only the metals,
but also the majority of the \HI\ gas that is confined within a small
cloud. \citet{Schaye07} posit a scenario where tiny, very metal rich
clumps populate \HI\ clouds in the IGM. In this scenario, these small
metal clumps are responsible for the metals seen in high
\NHI\ absorbers, which can have much larger sizes than the
clumps. However, there is no evidence that this is the case for the
absorbers we see: the \HI\ component velocities are extremely
well-aligned with the low ions (see Fig.~\ref{f_abs}), and the
photoionization models effectively reproduce the observed column
densities of both metals and \HI.

\subsection{Requirements for numerical simulations of the CGM to 
resolve these clouds}

\label{f_sim}

What are the requirements to resolve the hydrodynamic instabilities
affecting these clouds? The situation is comparable to the `blob test'
performed by \citet{Agertz07}, where a test case of a dense spherical
clump of gas travelling through a diffuse medium is considered using
both adaptive mesh refinement (AMR) and smoothed particle
hydrodynamics (SPH) simulations. Agertz et al. find that 7 adaptive
mesh resolution elements per cloud radius are needed to begin
resolving the hydrodynamic instabilities which destroy the clump. We
adopt an optimistic minimum requirement of 3 resolution elements per
clump radius. Assuming a radius of 400~proper pc---at the upper end of
our measured sizes---this implies a minimum cell size of 140
pc. Therefore AMR simulations must refine to this scale in the CGM at
densities of $10^{-3}$--$10^{-2}$~\cmmm. In existing AMR simulations
such small refinement scales can be reached, but only for the highest
density regions with $\sim1$~\cmmm.

For SPH simulations, the resolution requirements are best expressed in
terms of particle mass. \citet{Agertz07} show the number of
independent resolution elements per clump radius is $n_{k} = 1/2
(N_{\rm p} / n_{\rm smooth})^{1/3}$, where $N_{\rm p}$ is the number
of particles in the clump and smoothing is performed over $n_{\rm
  smooth}$ nearest neighbours. Taking typical values of $n_{\rm
  smooth}=32$ and $n_\mathrm{k}=3$ requires $N_\mathrm{p}=6912$
particles in the clump. We estimate the mass of the absorbing clumps
assuming spherical clouds with a mass given by
\begin{equation}
\begin{split}
M_{\rm clump} & = 4/3\, \pi\, R_{\rm clump}^3\,\nH\, \mu m_{\rm p}  \\
            &  = 2.8 \times 10^4 \msun \left( \frac{R_{\rm clump}}{0.4\,{\rm kpc}} \right)^3 \left( \frac{\nH}{10^{-2.5}\,\cmmm}\right)
\end{split}
\end{equation}
where $\mu m_\mathrm{p} = 1.4 m_\mathrm{p}$ is the mass per hydrogen
atom. The masses we measure are in Table~\ref{t_par}.  They imply that
the required SPH particle mass is $<4\,$\msun, orders of magnitude
smaller than the particle mass of even the most recent zoomed SPH
simulations of single galaxies: ERIS2 \citep[][$\sim 2 \times
  10^4$\msun]{Shen13} and FIRE \citep[][$\sim5 \times
  10^3$\msun]{Hopkins14}. Simulations which aim to simulate
representative volumes of the Universe, such as Illustris,
\citep[][with a mass resolution of $\sim 10^6$\msun]{Vogelsberger14}
are even further from being able to resolve these clumps. We stress
that this is a minimum required mass resolution. We have assumed a
clump radius 400~pc when the size of many low-ion absorbers may be
$<100$~pc, and that only 3 resolution elements per clump radius are
required to correctly resolve hydrodynamic instabilities.

We conclude that to date, no single-galaxy or cosmological simulation
has been performed at a sufficient resolution to predict the
properties of low-ion gas in the CGM. The best approach for future
simulations may be to abandon the attempt to resolve these gas clumps
altogether. Instead a sub-grid model for their behaviour could be
introduced, in the same way that sub-grid models are used to treat
star formation and galactic-scale outflows in these simulations.

\section{Summary}
\label{s_summary}

We have serendipitously discovered a faint, $0.2\,L^*$, galaxy at
$z=2.466$ that is at a small impact parameter ($50$~kpc) from a
background QSO. Based on similarities to galaxies selected as
\lya\ emitters, it likely has a halo mass $\sim 10^{11.4}\msun$ and
stellar mass $< 10^{10}\msun$.

A very high quality UVES spectrum of the QSO reveals a partial Lyman
limit system at the galaxy redshift. The absorber shows low- and
high-ionization potential transitions in several components spanning a
velocity width of $436~\kms$. As this galaxy-absorber pair was not
selected based on the absorber properties, it is likely that the
covering fraction of such absorbers around similar $z=2.5$ galaxies is
significant. Our results can be summarised as follows:
\begin{enumerate}
\item Using a new method to marginalise over uncertainties in the
  shape of the incident UV field, we measure the density and
  metallicity of each absorbing component. A single photoionized gas
  phase can explain all of the low-ionization absorption we see. The
  gas has a metallicity, $Z=0.1$--$0.6~Z_\odot$, similar
  to the ISM of $z\sim2.5$ galaxies. The gas temperature is cool
  ($10^4$~K) with densities $10^{-3}$ to $10^{-2}$~\cmmm.
\item Infalling or tidally stripped gas is unlikely to explain all of
  the absorption, due to the large velocity width and large covering
  fraction of the gas. Our favoured model is an outflowing shell of
  gas. If most of the gas we see is in an outflow, the mass outflow
  rate implied is $\sim 5\msunyr$ with a factor of $\sim4$
  uncertainty due to an unknown outflow geometry. This is similar to
  outflow rates measured around galaxies with halo mass
  $10^{11}$--$10^{12}\msun$ at redshifts $z\sim0.7$ using
  `down-the-barrel' \MgII\ absorption against background galaxies.
\item \OVI\ gas is present with a broader linewidth than expected from
  the low-ions, and a higher \NOVI\ than predicted by our
  photoionization models. It must be produced by a different gas phase
  that is physically associated with the photoionized gas. This phase
  may be explained by a warmer, possibly radiatively cooling,
  envelope around the cooler, photoionized gas.
\item The total gas mass in the $10^4$~K, photoionized phase and the
  different phase traced by \OVI\ is comparable. Together they imply a
  baryonic mass $>0.8\times10^{9}\msun$ in the galaxy's CGM. The mass
  could be much higher if the halo extends further than 50~kpc, the
  covering fraction of gas is larger than 50\%, or if we use less
  conservative assumptions about the \OVI\ fraction and metallicity.
\item The photoionized clumps have sizes ranging from $100$--$500$
  pc. The uncertainties in these size estimates take into account
  uncertainties in the shape and normalisation of the UV background
  radiation. Kelvin-Helmholtz instabilities imply that the lifetime of
  clumps this size moving through a more diffuse halo are very
  short. Therefore, we suggest that there must be a mechanism present
  to make the clumps resistant to these instabilities, such as a warm
  gas envelope surrounding them. The \OVI\ absorption may be produced
  by such an envelope.
\item The small clump size makes them very difficult to resolve in
  numerical simulations. No single-galaxy or cosmological simulation
  performed to date has the resolution necessary to correctly treat
  the movement of the clumps through the CGM. We suggest that a CGM
  sub-grid model capturing the relevant physics may be necessary for
  future simulations.
\item In our outflow model, the wind velocities required to explain
  the kinematics exceed the escape velocity of a $10^{11}\msun$
  halo. In this case the metal-enriched gas will escape into the IGM.
\end{enumerate}
This final result hinges on the galaxy having a halo mass lower than is
typically measured for brighter $z\sim 2.5$ galaxies. We argue this is
likely to be the case based on the large \lya\ equivalent width and
faint rest-frame UV continuum. However, deep IR imaging would allow
more robust SED modelling of the galaxy to give a more precise
measurement of the stellar and halo masses.

Finally, the new integral field unit now available on the ESO Very
Large Telescope, MUSE, has the capablity to find similar weak emission
line galaxies around QSO sightlines at higher redshifts. We expect
MUSE will be able to efficiently assemble a large number of similar
systems over the redshift range $3$--$5$.

\vspace{0.5cm}

We thank the referee for their helpful comments which improved the
paper. Yujin Yang provided helpful correspondence, and Elisabeta Da
Cuhna generously provided electronic versions of the tables in her
paper. We also thank Richard Bielby, Charles Finn, Simon L. Morris and
Kate Rubin for their comments on a draft of this paper. This work was
based on observations carried out at the European Southern Observatory
(ESO), under programs 166.A-0106, 185.A-0745 and 091.A-0698. We
particularly thank the Paranal Observatory support staff for their
help with the visitor-mode observations taken for program 091.A-0698.
Our analysis made use of \textsc{astropy}
\citep{Astropy13}\footnote{\url{http://www.astropy.org}},
\textsc{xidl}\footnote{\url{http://www.ucolick.org/~xavier/IDL}},
\textsc{DS9}\footnote{\url{http://ds9.si.edu}} and the NASA/IPAC
Extragalactic Database
(NED)\footnote{\url{http://ned.ipac.caltech.edu}}. Plots were made
with \textsc{matplotlib} (Hunter et
al. 2007)\footnote{\url{http://matplotlib.org}}
and \textsc{aplpy}\footnote{\url{http://aplpy.github.com}}. NC and MM
thank the Australian Research Council for \textsl{Discovery Project}
grant DP130100568 which supported this work. MF acknowledges support
by the Science and Technology Facilities Council [grant number
  ST/L00075X/1].

\footnotesize{

}

\appendix
\section{Virial radius, temperature and velocity}

We use the analytic expression for the virial overdensity $\delta_{\rm
  vir} \equiv \rho_{\rm vir}/ \rho_{\rm crit}$ from
\citet{Bryan98}. Then we use the following expressions to find the
virial radius, circular velocity, and temperature:
\begin{equation*}
  r_{\rm vir} = \left( \frac{3 M_{\rm halo}}{4 \pi \rho_{\rm
      vir}} \right)^{1/3}
\end{equation*}
\begin{equation*}
  v_{\rm circ} = \sqrt{G M_{\rm halo} / r_{\rm vir}}
\end{equation*}
\begin{equation*}
  T_{\rm vir} = \frac{\mu m_{\rm p} v_{\rm circ}^2}{2 k}
\end{equation*}
Here $\mu m_{\rm p}$ is the mean particle mass and we assume
$\mu=0.59$, for a primordial ionized gas. $G$ is the gravitational
constant, $k$ is Boltzmann's constant, and $M_{\rm halo}$ is the
galaxy dark matter halo mass.

\section{Estimating physical parameters from \textsc{cloudy} modelling including an uncertainty in 
the shape of the ionizing radiation.}
\label{a_Cloudy}
\textsc{cloudy} models predict the column densities, $N_i$, for $\mathcal{N}$ species as
a function of the hydrogen volume density \nH, metallicity $Z$ and the
neutral hydrogen column density \NHI\ of an absorbing gas cloud,
assuming some functional form for the radiation field, $F_\nu$, at the
cloud surface. Thus
\begin{equation}
N_i = N_i(\nH, Z, N_{\rm HI})
\end{equation}
\noindent where $i =$ \CII, \SiII, \MgII, \OI, \CIII, \SiIII, ... with
a total of $\mathcal{N}$ species. The hydrogen volume density \nH\ together with
the ionizing spectrum can be used to find the ionization parameter
$U$, often quoted by \textsc{cloudy} analyses in the literature:
\begin{equation}
\label{e_U}
U \equiv \frac{n_\gamma}{\nH}  = \frac{\Phi_\gamma}{\nH c}  = \frac{1}{\nH c} \int^\infty_{\nu_{\rm LL}} \frac{F_\nu}{h\nu} d\nu
\end{equation}
Here $n_\gamma$ is the density of photons able to ionize hydrogen,
$\Phi_\gamma$ is the corresponding photon flux, $\nu$ is the
frequency, $c$ is the speed of light, $\nu_{\rm LL}$ is the frequency
corresponding to the hydrogen Lyman limit, and $F_{\nu}$ has units
erg~s$^{-1}$~cm$^{-2}$~Hz$^{-1}$. Thus the ionization parameter is the
ratio of the density of photons able to ionize neutral hydrogen to the
hydrogen density at the face of the cloud.  Often \NHI\ is assumed to
be precisely known for the cloud, and a 2 dimensional grid of
\textsc{cloudy} models with a single \NHI\ stopping criterion is used
to find the most likely values of $Z$ and $\nH$ given the data.  The
likelihood $\mathcal{L}$ of the observed data given an input $Z$ and
$\nH$ is then
\begin{equation}
\mathcal{L}(Z,\nH) = \prod\limits_{i\,=\,1}^\mathcal{N}\ell_i(Z,\nH)
\end{equation}
\noindent where $\ell_i(Z,U)$ is the probability that the $i$th
predicted column density is drawn from the distribution given by the
observed value and its uncertainty. The observed column densities are
either measurements with a lower and upper uncertainty, upper limits,
or lower limits.  As the likelihoods can be extremely small, we
instead calculate the natural logarithm of the likelihood to avoid
computational problems related to the representation of floating-point
numbers. Therefore:
\begin{equation}
\ln \mathcal{L}(Z,\nH) = \sum\limits_{i\,=\,1}^\mathcal{N} \ln(\, \ell_i(Z,\nH)\, )
\end{equation}

\begin{figure}
\includegraphics[width=\linewidth]{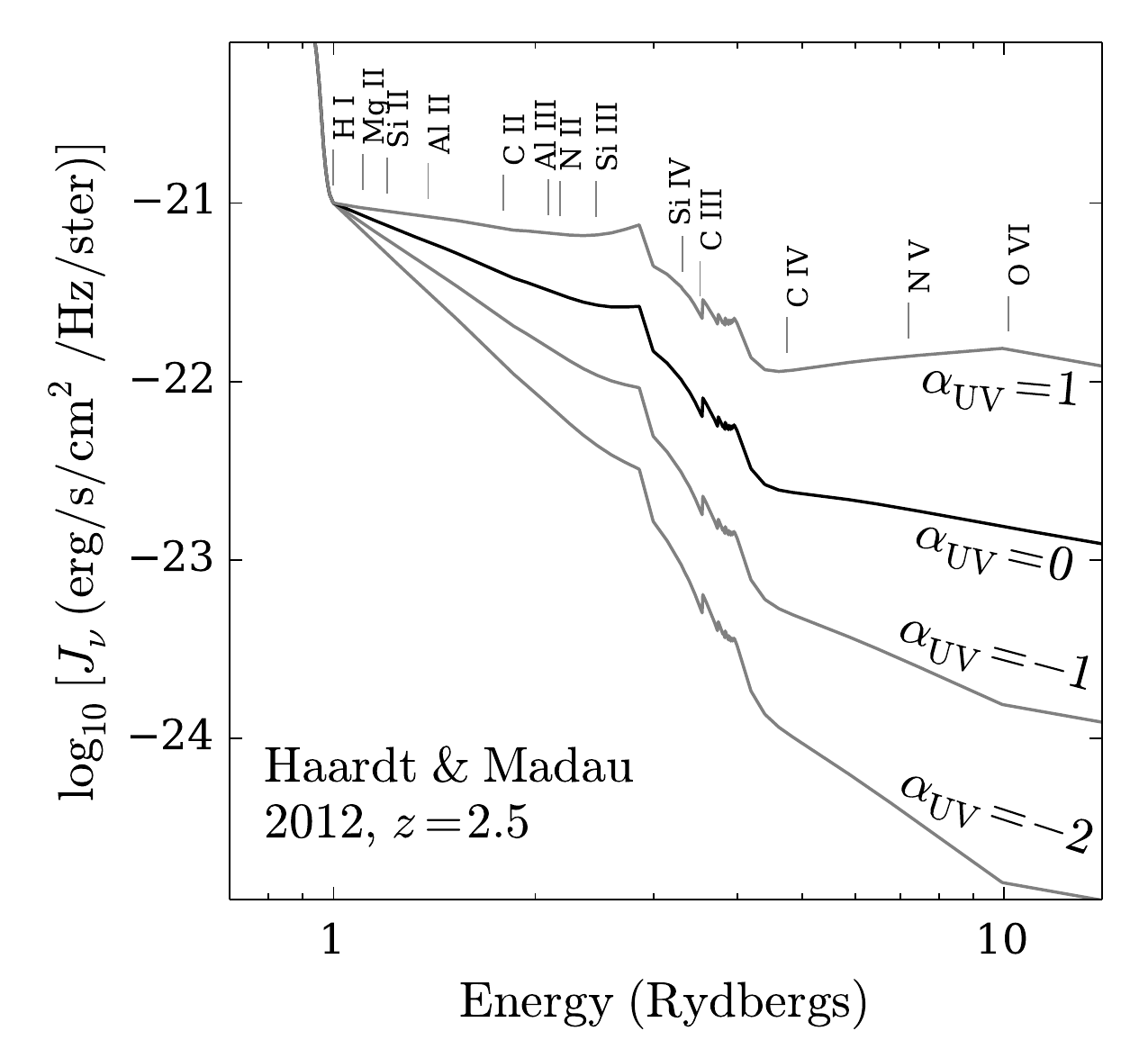}
\caption{\label{f_aUV1a}  The effect of different
  $\alpha_\mathrm{UV}$ on the shape of the integrated QSO \& galaxy
  background radiation. $\alpha_\mathrm{UV}=0$ gives the fiducial HM12
  background spectrum at the redshift of the
  absorber. $\alpha_\mathrm{UV}>0$ makes the slope more shallow
  (giving a harder spectrum) while $\alpha_\mathrm{UV}<0$ makes it
  steeper (softer). The ionization potential of the observed
  transitions is also shown.}
\end{figure}

\begin{figure}
\includegraphics[width=\linewidth]{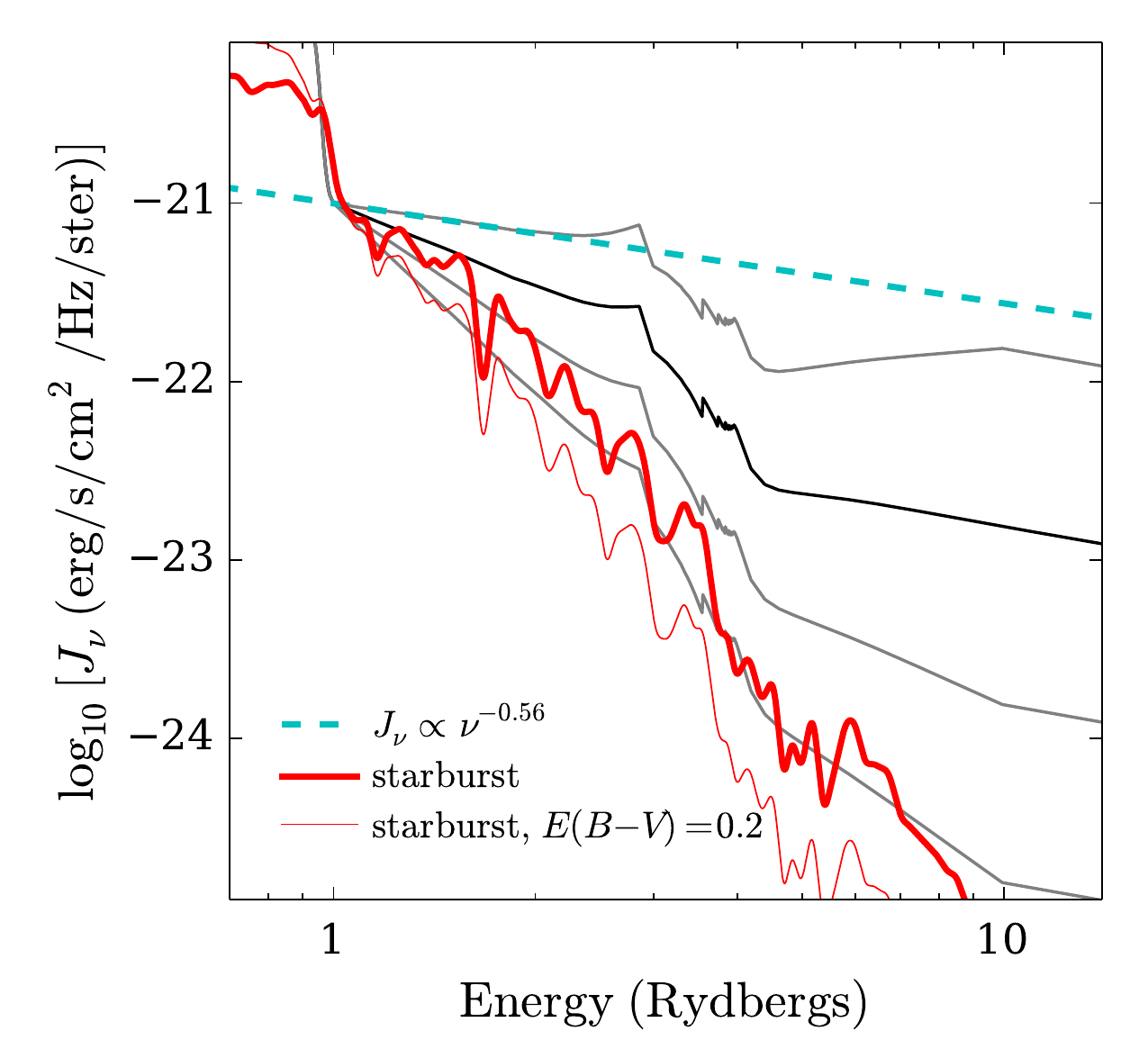}
\caption{\label{f_aUV1b} Comparison between QSO-dominated and
  galaxy-dominated radiation fields, and our parametrisation. The
  dashed line shows a hard QSO spectrum with spectral index
  $-1.5$. The thick red line shows a smoothed spectrum of starburst
  galaxy generated using Starburst99. The thinner red line below shows
  the same spectrum with the extinction from \citep{Calzetti00}
  applied, assuming $E(B-V)=0.2$, typical of $z=2.5$ Lyman break
  galaxies \citep{Reddy08}. The range of $\alpha_{\rm UV}$ we consider
  covers both these extreme cases. The black and gray lines are the
  same as those shown in Fig.~\ref{f_aUV1a}.}
\end{figure}

\begin{figure}
\includegraphics[width=\linewidth]{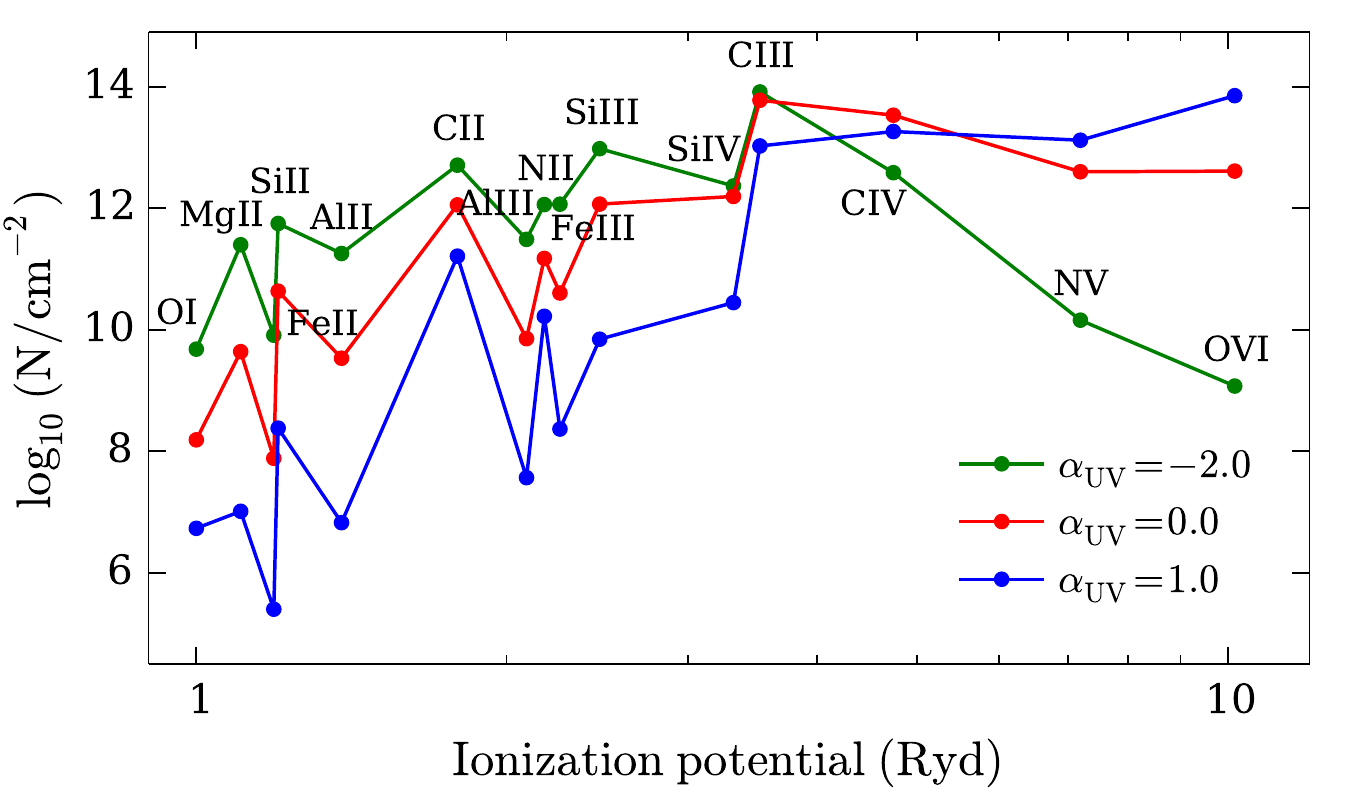}
\caption{\label{f_aUV2} The effect of different $\alpha_\mathrm{UV}$
  on predicted column densities. The predicted column densities are
  plotted as a function of the species ionization potential for a
  fixed metallicity, \nH\ and \NHI, only varying
  $\alpha_\mathrm{UV}$. As $\alpha_\mathrm{UV}$ becomes more negative
  (giving a softer ionizing spectrum), the column density of low
  ionization potential species (left) increases with respect to higher
  ionization potential species (right).}
\end{figure}

In this work we include two more parameters in addition to $Z$ and
\nH: a UV shape parameter $\alpha_\mathrm{UV}$, and \NHI, for
absorbers where it is not well determined.  The shape parameter
modifies the reference HM12 UV field in the following way:
\begin{multline}
\log_{10} F_{\nu}(E) = \\
 \begin{cases} 
      H(E)   & E < E_0  \\
      H(E)~\alpha_\mathrm{UV}\log_{10}(E/E_0)  &  E_0 \le E \le E_1 \\
      H(E) + H(E_1)\,\alpha_\mathrm{UV}\log_{10}(E_1/E_0) - H(E_1)  &  E > E_1 
   \end{cases}
\end{multline}
\noindent where $H(E)$ is $\log_{10}$ of the HM12 field, and we
set $E_0=1$~Ryd and $E_1=10$~Ryd to cover the ionization energies of
the observed transitions.  Figures~\ref{f_aUV1a} \& \ref{f_aUV1b} show
how different values of $\alpha_\mathrm{UV}$ affect the UV field
incident on the absorber, and how the range of slopes generated by our
parametrisation compares to the spectra of QSOs and starburst
galaxies. $\alpha_\mathrm{UV}=0$ leaves the default HM12 background
unchanged; $\alpha_\mathrm{UV}<0$ produces steeper, increasingly
negative power-law slopes with respect to the default (a softer
ionizing spectrum), and $\alpha_\mathrm{UV} > 0$ results in a
shallower, less negative slope (a harder ionizing spectrum).
Fig.~\ref{f_aUV2} shows how the predicted column densities change for
different values of $\alpha_\mathrm{UV}$.  

This is a simple parametrization of the spectral shape. A model for
the UV background with three or more free parameters has also been
proposed by \citet{Fechner11} and \citet{Agafonova07}. However, we
prefer a one-parameter model for two reasons: first, the absorbers we
model do not show enough metal transitions to effectively constrain
more than one parameter, and second, it is computationally simpler to
evaluate. Fig.~\ref{f_aUV1b} shows how this parametrisation compares
to the largest and smallest slopes expected for the ionizing spectrum:
$\alpha_\mathrm{UV}= -2$ gives a spectral shape close to that expected
for a starburst galaxy with dust extinction, whereas
$\alpha_\mathrm{UV}=1$ corresponds to the largest power law slope
observed in spectra of QSOs in the 1 to 4 Ryd range
\citep{Scott04,Shull12}. Therefore our one-parameter model captures
the range of uncertainty in the shape of the ionizing spectrum
relevant to the 1--10 Ryd region.


\begin{figure}
\includegraphics[width=1.04\linewidth]{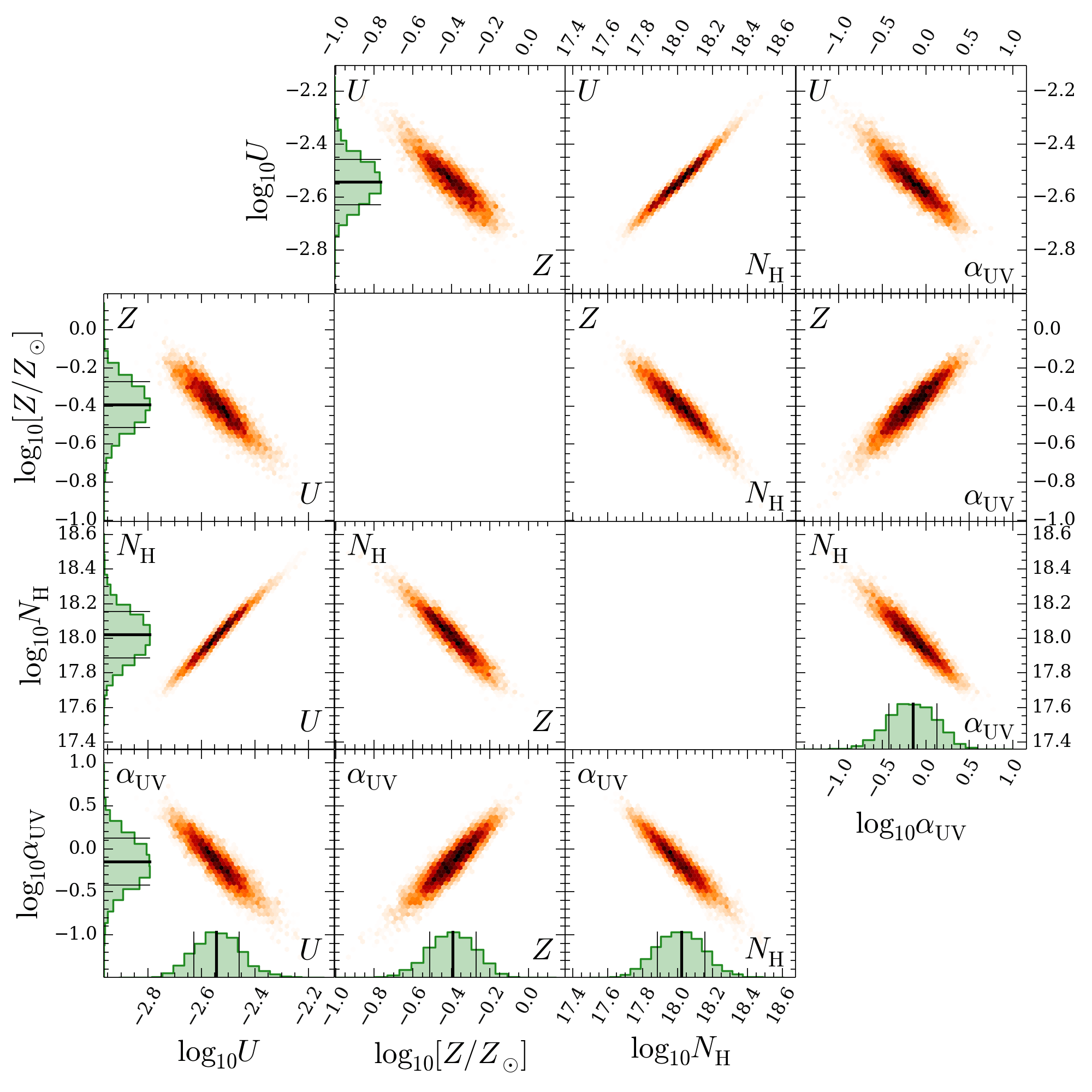}
\caption{\label{f_pos2} Posterior distributions for component 2. See the
  Fig.~\ref{f_pos1} caption for details.}
\end{figure}

\begin{figure}
\includegraphics[width=1.04\linewidth]{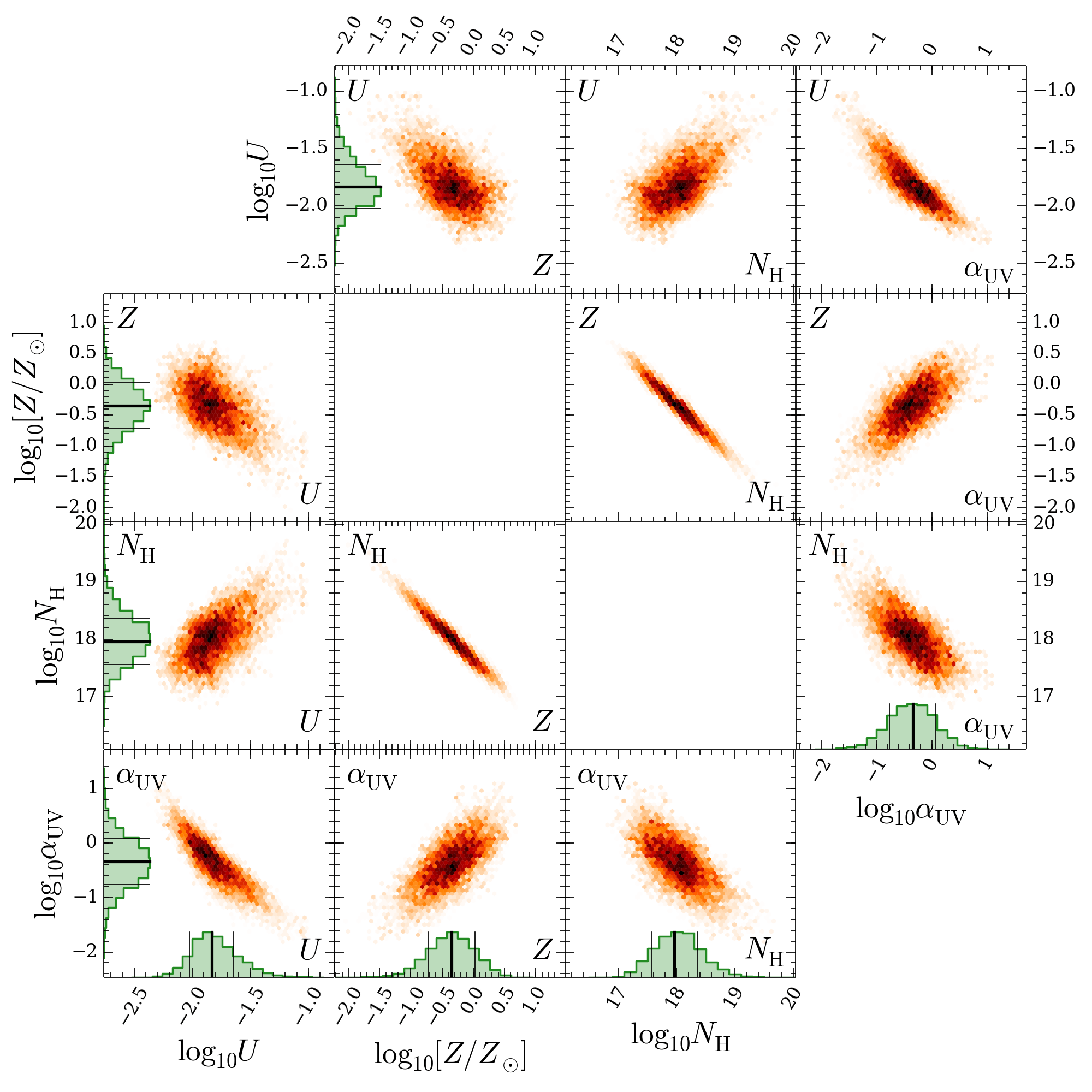}
\caption{\label{f_pos3} Posterior distributions for component 3. See the
  Fig.~\ref{f_pos1} caption for details.}
\end{figure}

\begin{figure}
\includegraphics[width=1.04\linewidth]{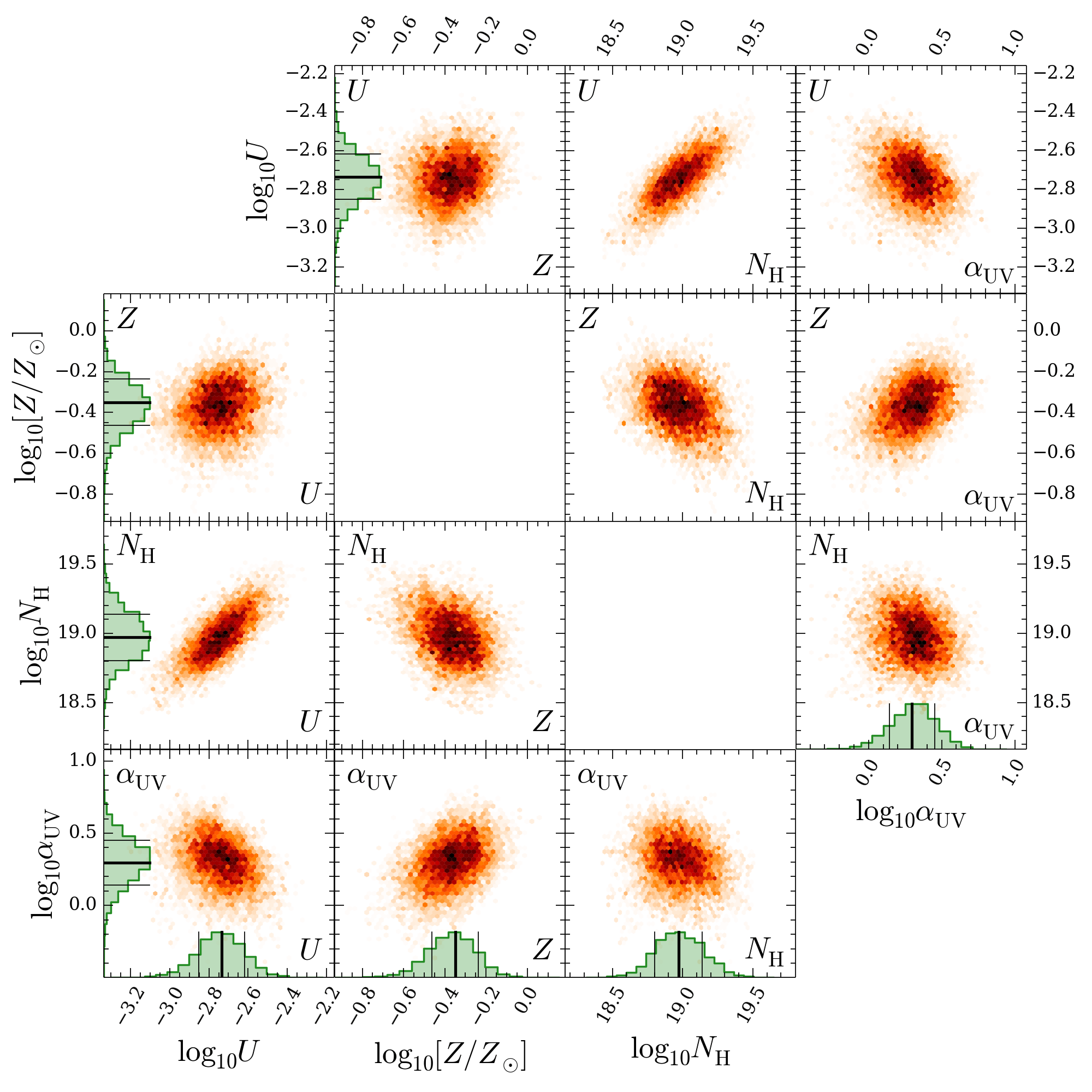}
\caption{\label{f_pos4} Posterior distributions for component 4. See the
  Fig.~\ref{f_pos1} caption for details.}
\end{figure}

\begin{figure}
\includegraphics[width=1.04\linewidth]{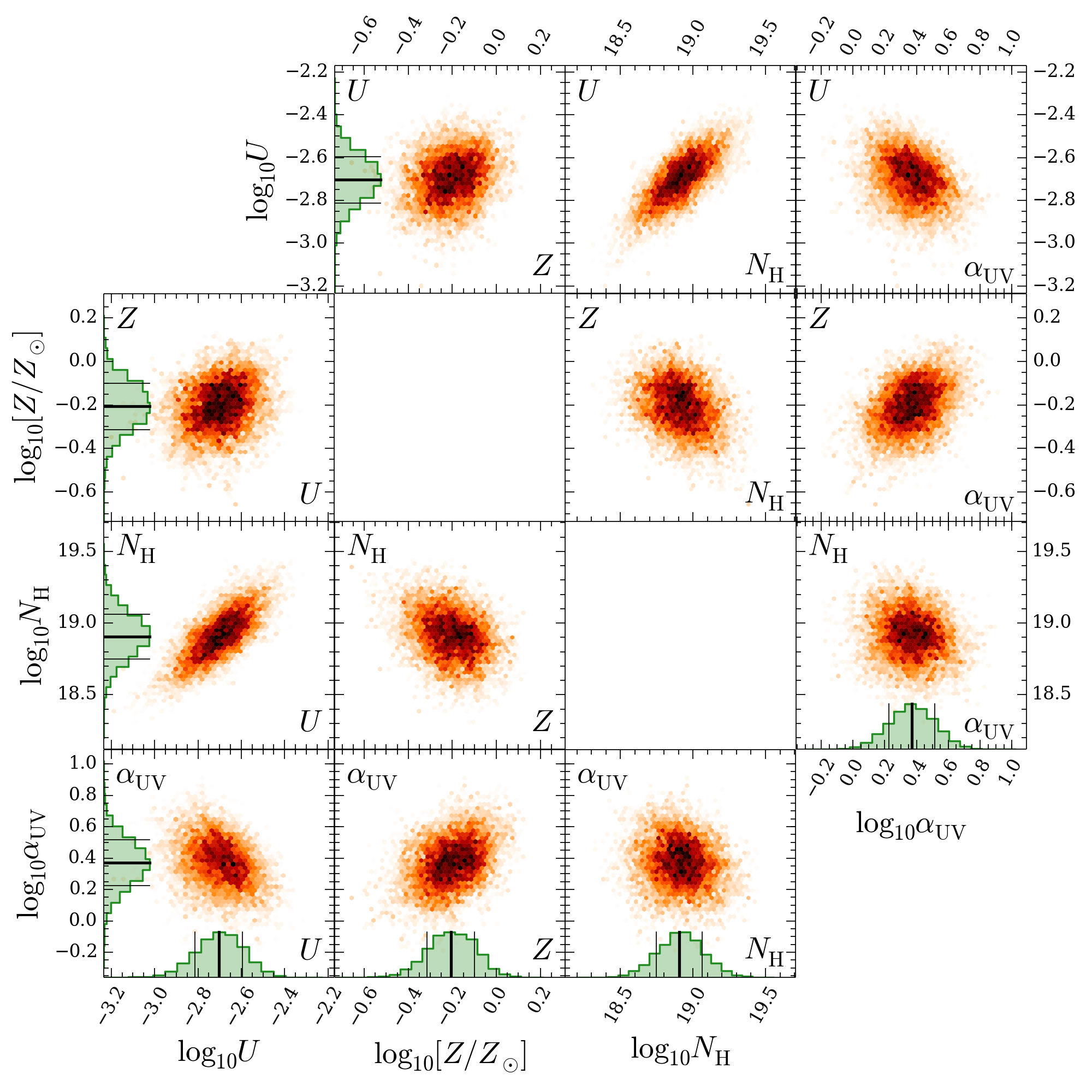}
\caption{\label{f_pos5} Posterior distributions for component 5. See the
  Fig.~\ref{f_pos1} caption for details.}
\end{figure}

\begin{figure}
\includegraphics[width=1.04\linewidth]{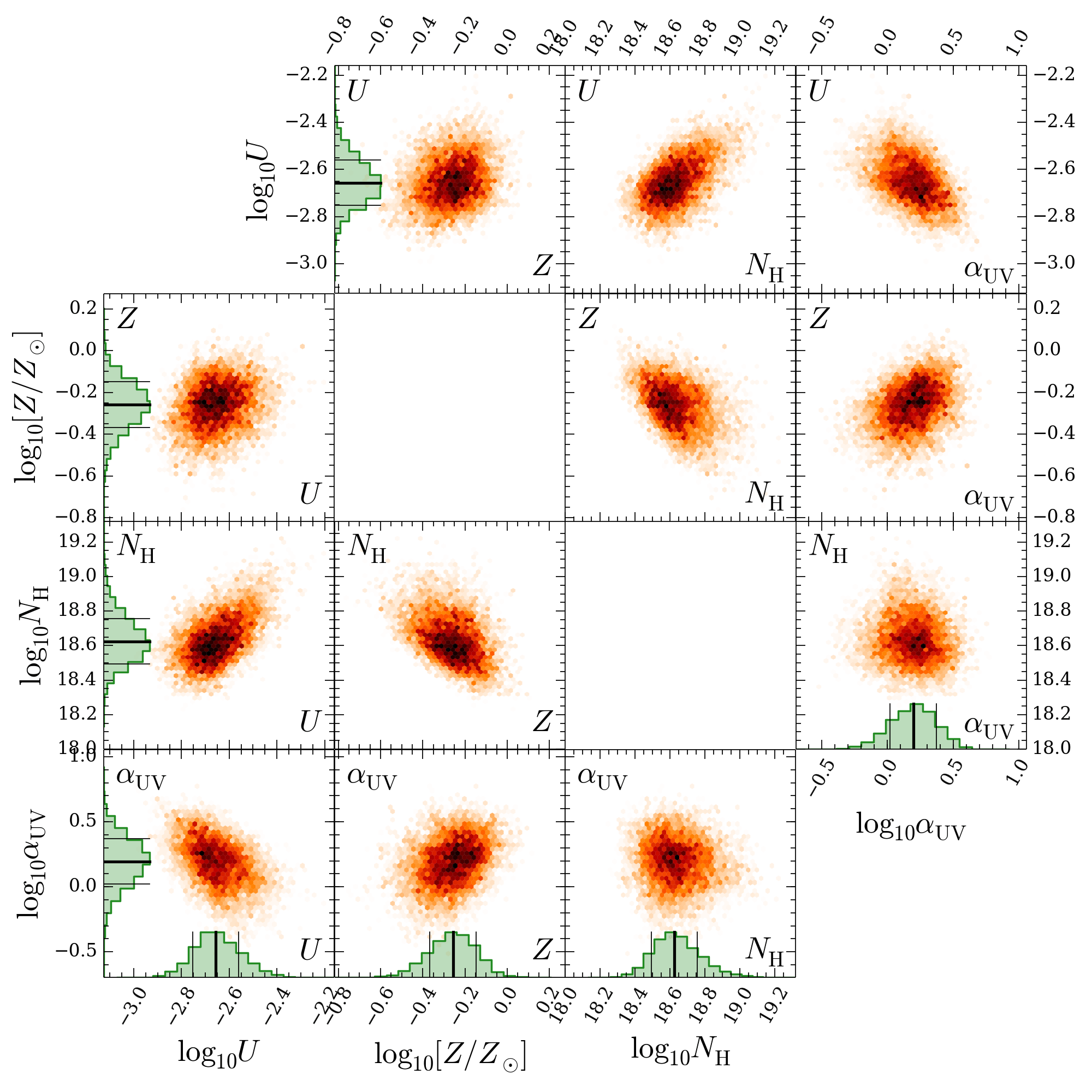}
\caption{\label{f_pos6} Posterior distributions for component 6. See the
  Fig.~\ref{f_pos1} caption for details.}
\end{figure}

\begin{figure}
\includegraphics[width=1.04\linewidth]{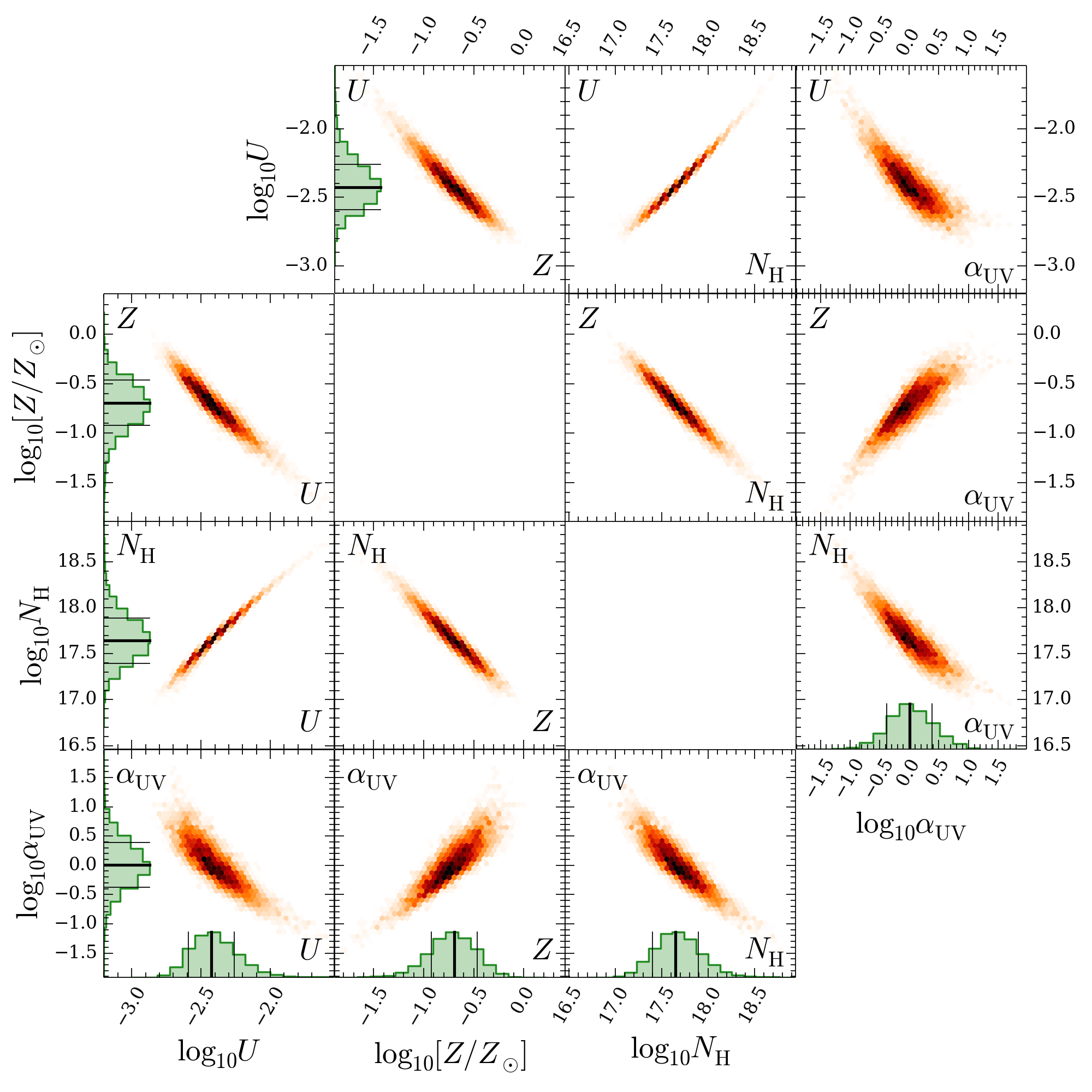}
\caption{\label{f_pos7} Posterior distributions for component 7. See the
  Fig.~\ref{f_pos1} caption for details.}
\end{figure}

\begin{figure}
\includegraphics[width=1.04\linewidth]{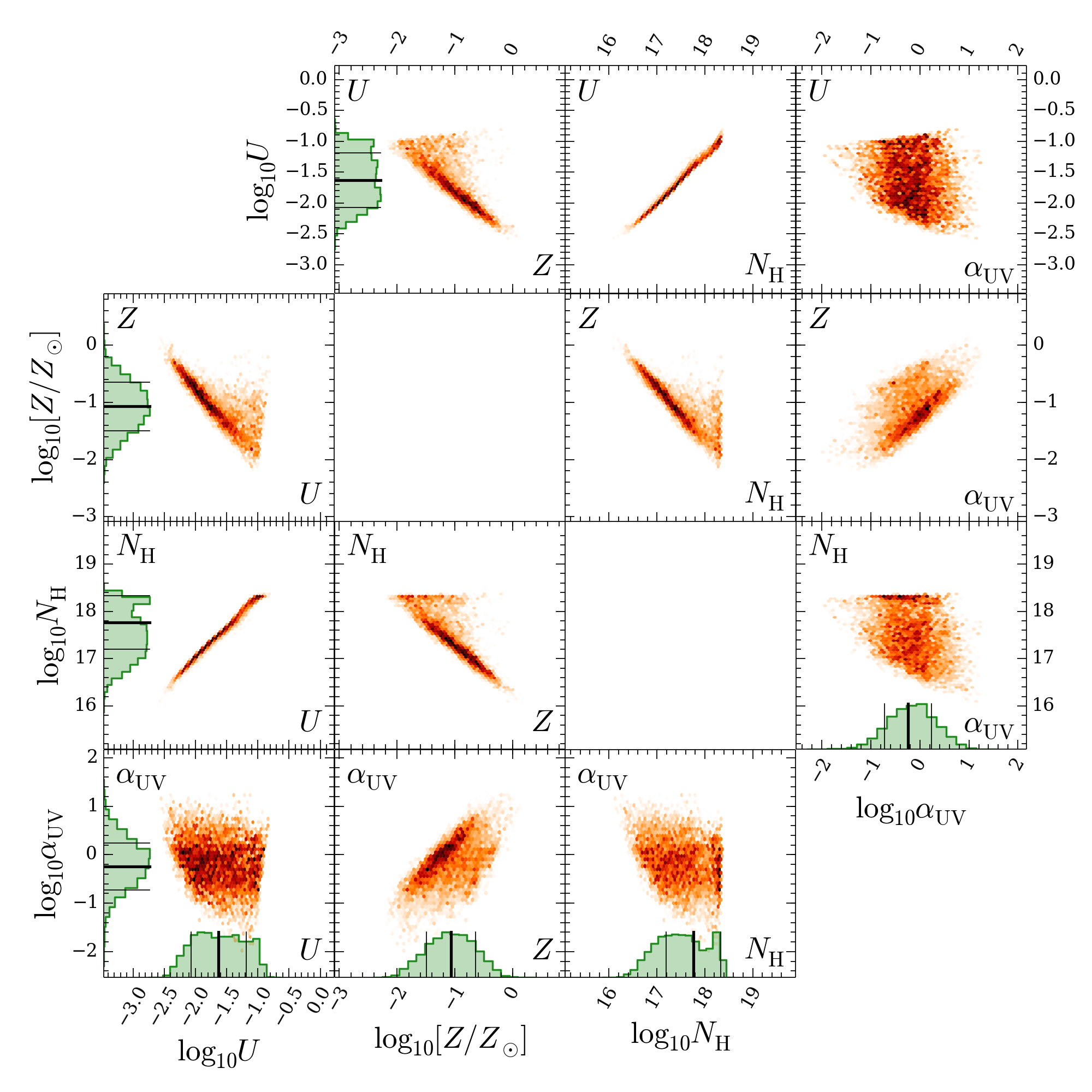}
\caption{\label{f_pos8} Posterior distributions for component 8.  See
  the Fig.~\ref{f_pos1} caption for details. Note that $U$ must be
  $<10^{-1}$ for this component, because higher $U$ implies a
  temperature $> 10^{4.5}$~K, which is ruled out by the
  \HI\ linewidth, $21.1\pm0.3$~\kms.}
\end{figure}

\begin{figure}
\includegraphics[width=1.04\linewidth]{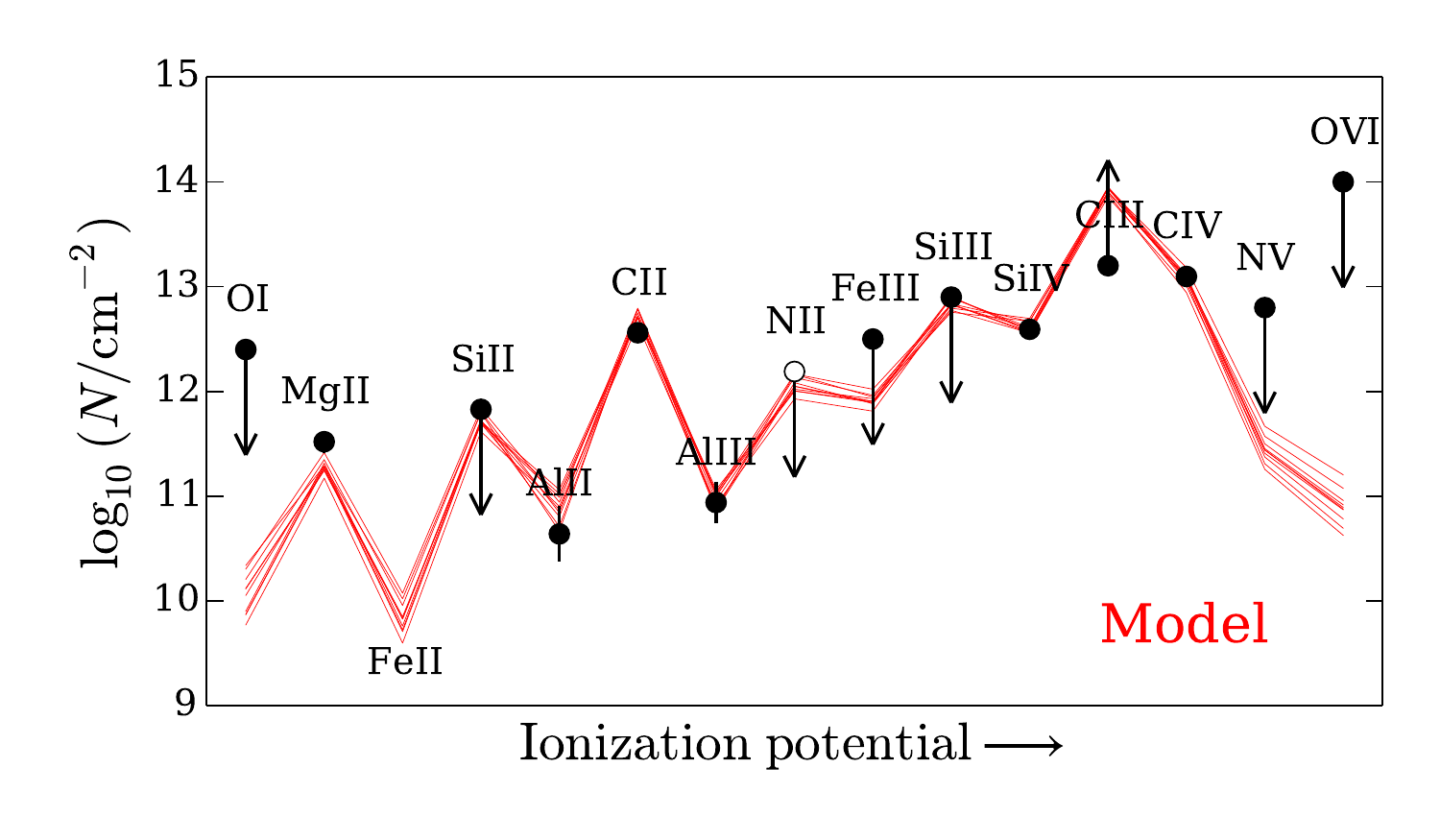}
\caption{\label{f_model2} Ten MCMC samples for component 2. See the Fig.\ref{f_model1} caption for details.}
\end{figure}

\begin{figure}
\includegraphics[width=1.04\linewidth]{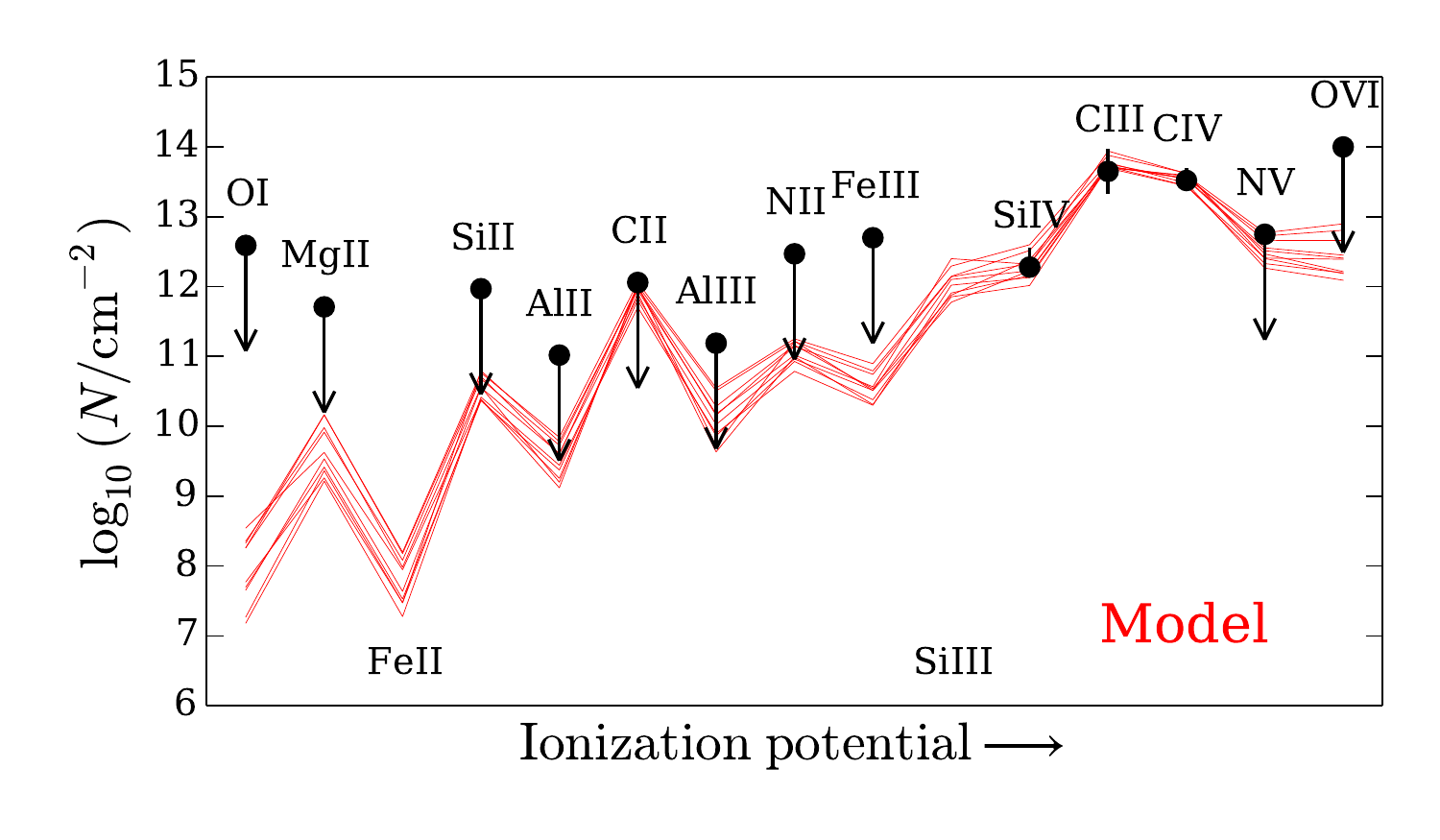}
\caption{\label{f_model3} Ten MCMC samples for component 3. See the Fig.\ref{f_model1} caption for details.}
\end{figure}

\begin{figure}
\includegraphics[width=1.04\linewidth]{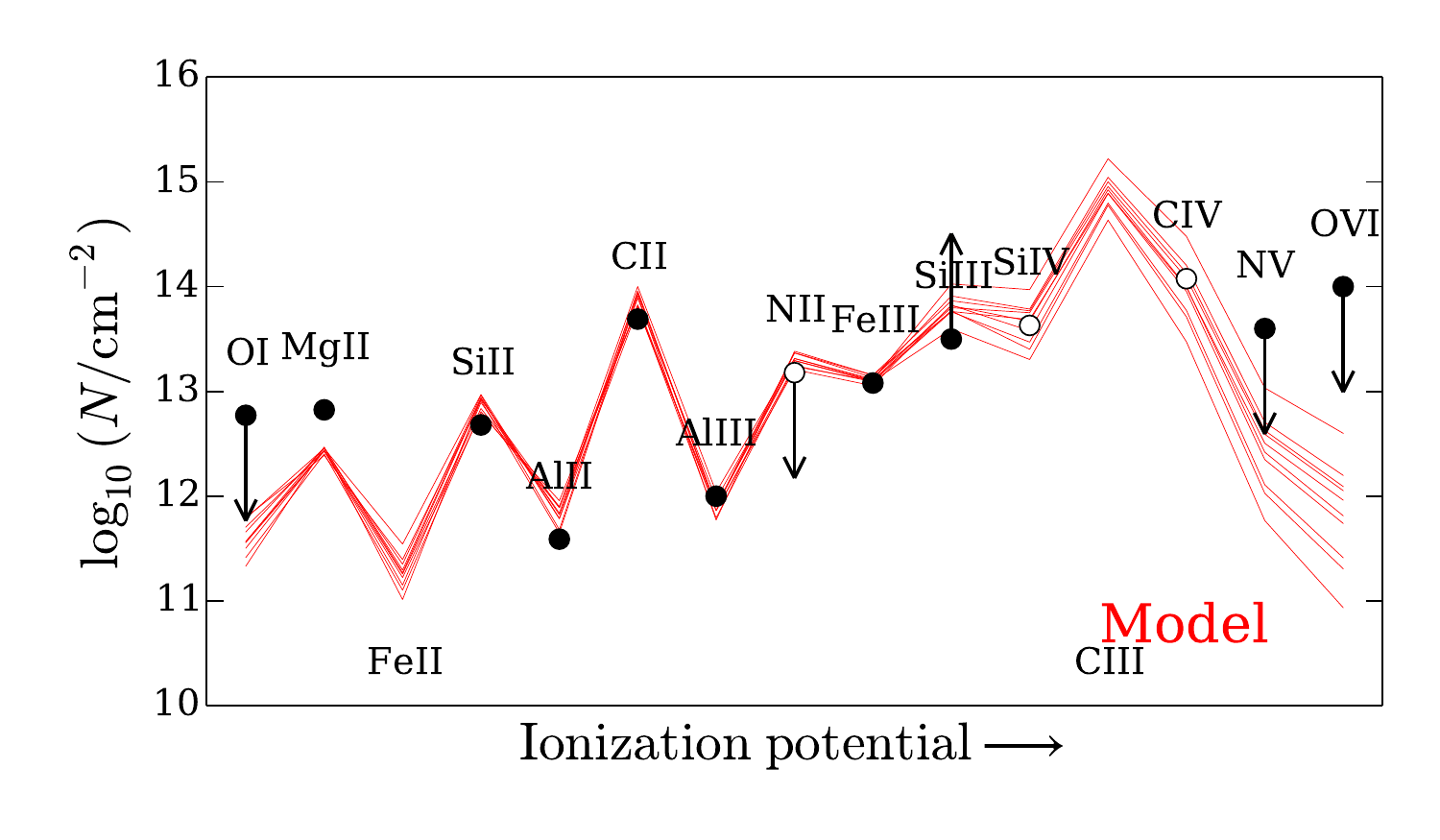}
\caption{\label{f_model4} Ten MCMC samples for component 4. See the
  Fig.\ref{f_model1} caption for details. \CIV\ and \SiIV\ were not
  used in parameter estimation in this component, as they are strongly
  blended with unrelated components.}
\end{figure}

\begin{figure}
\includegraphics[width=1.04\linewidth]{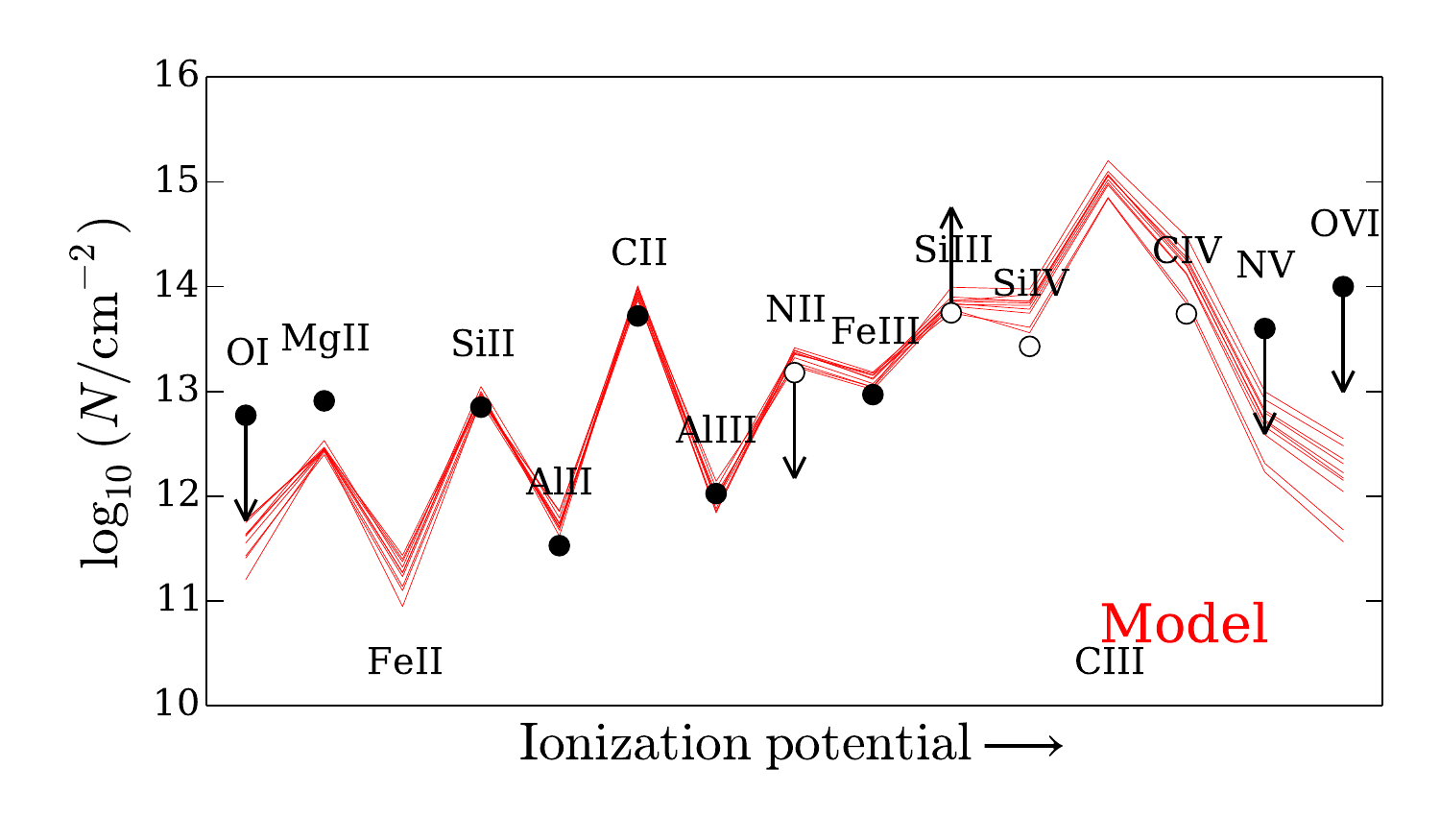}
\caption{\label{f_model5} Ten MCMC samples for component 5. See the
  Fig.\ref{f_model1} caption for details. \CIV\ and \SiIV\ were not
  used in parameter estimation in this component, as they are strongly
  blended with unrelated components.}
\end{figure}

\begin{figure}
\includegraphics[width=1.04\linewidth]{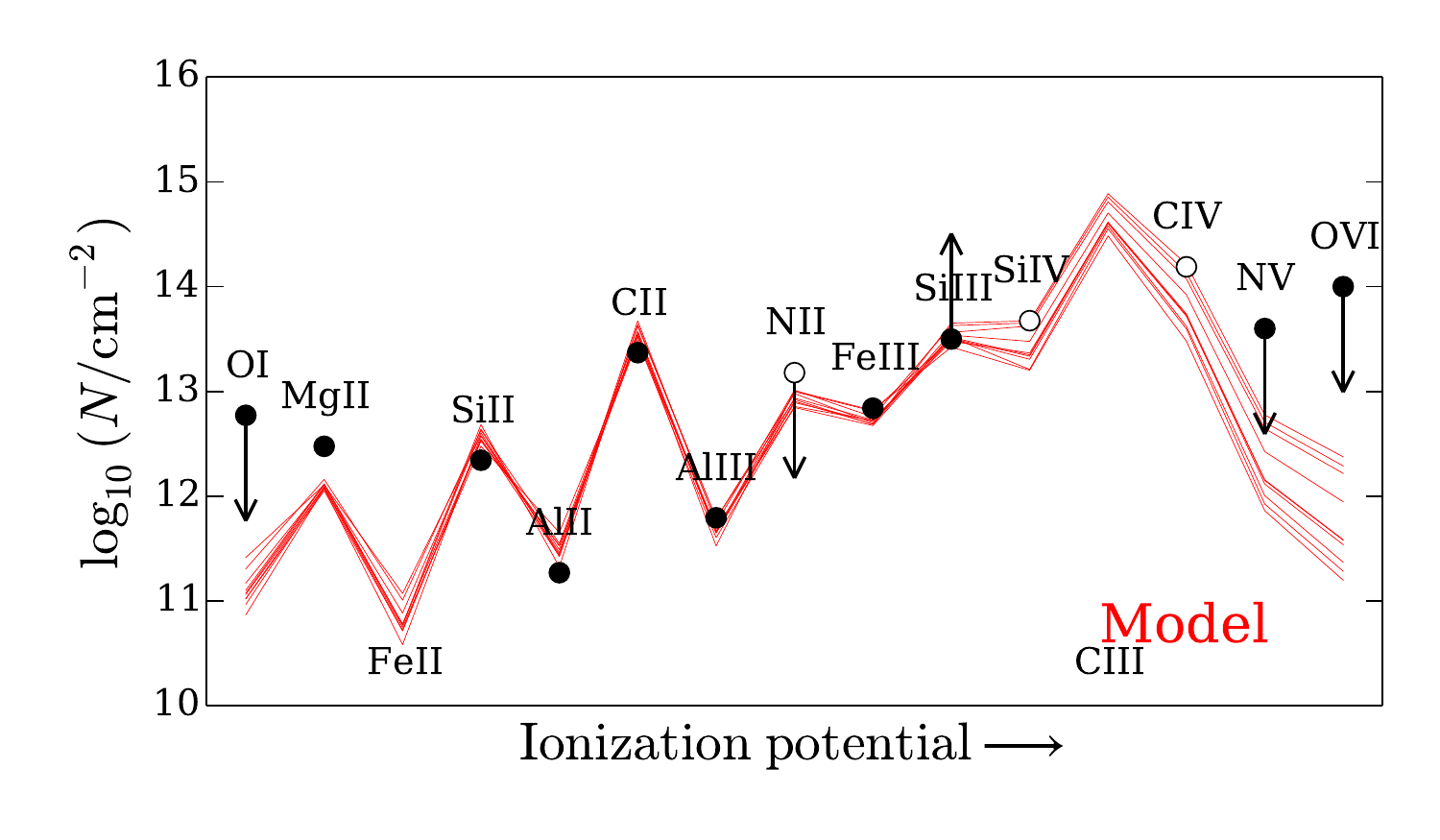}
\caption{\label{f_model6} Ten MCMC samples for component 6. See the
  Fig.\ref{f_model1} caption for details. \CIV\ and \SiIV\ were not
  used in parameter estimation in this component, as they are strongly
  blended with unrelated components.}
\end{figure}

\begin{figure}
\includegraphics[width=1.04\linewidth]{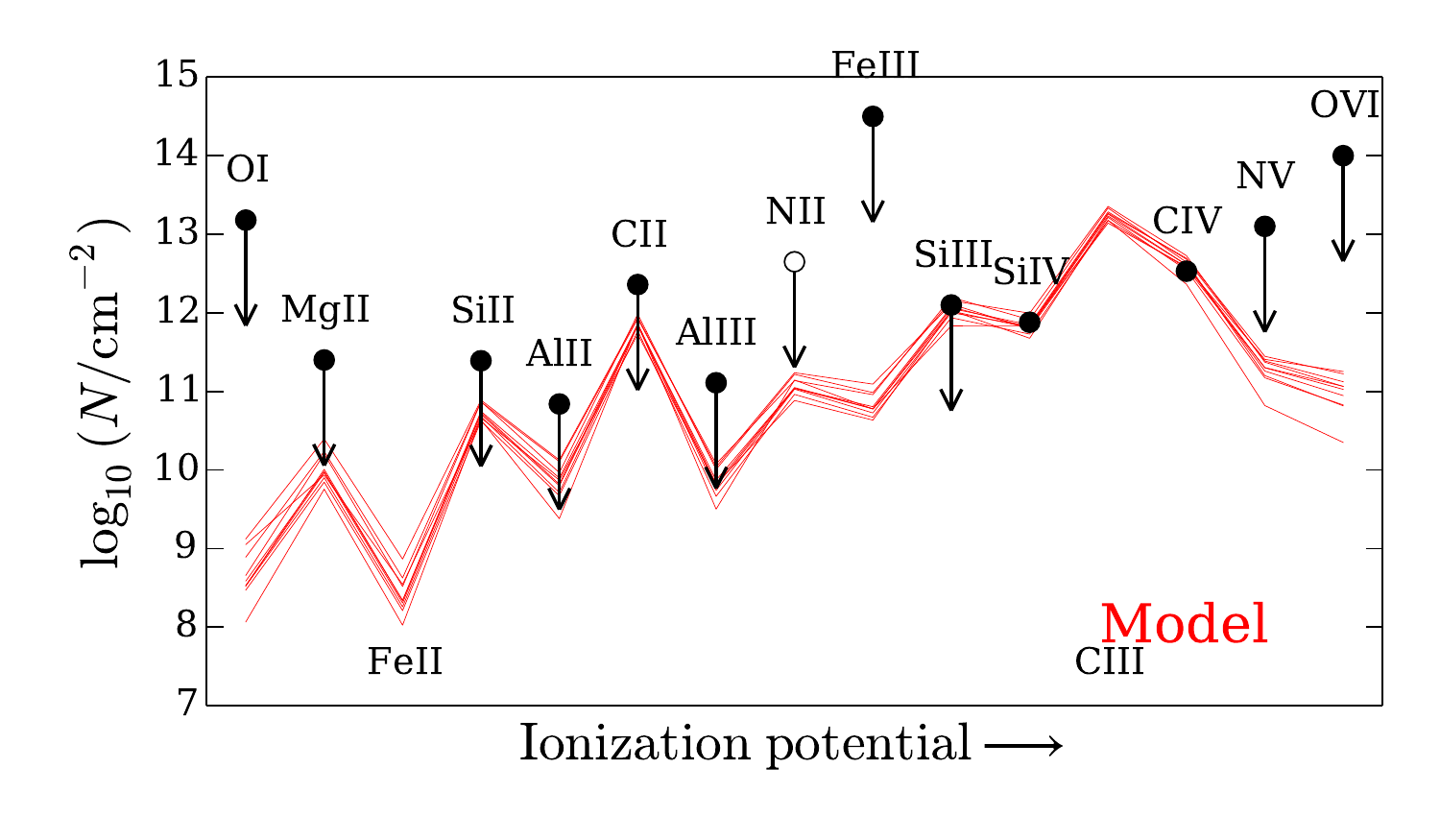}
\caption{\label{f_model7} Ten MCMC samples for component 7. See the Fig.\ref{f_model1} caption for details.}
\end{figure}

\begin{figure}
\includegraphics[width=1.04\linewidth]{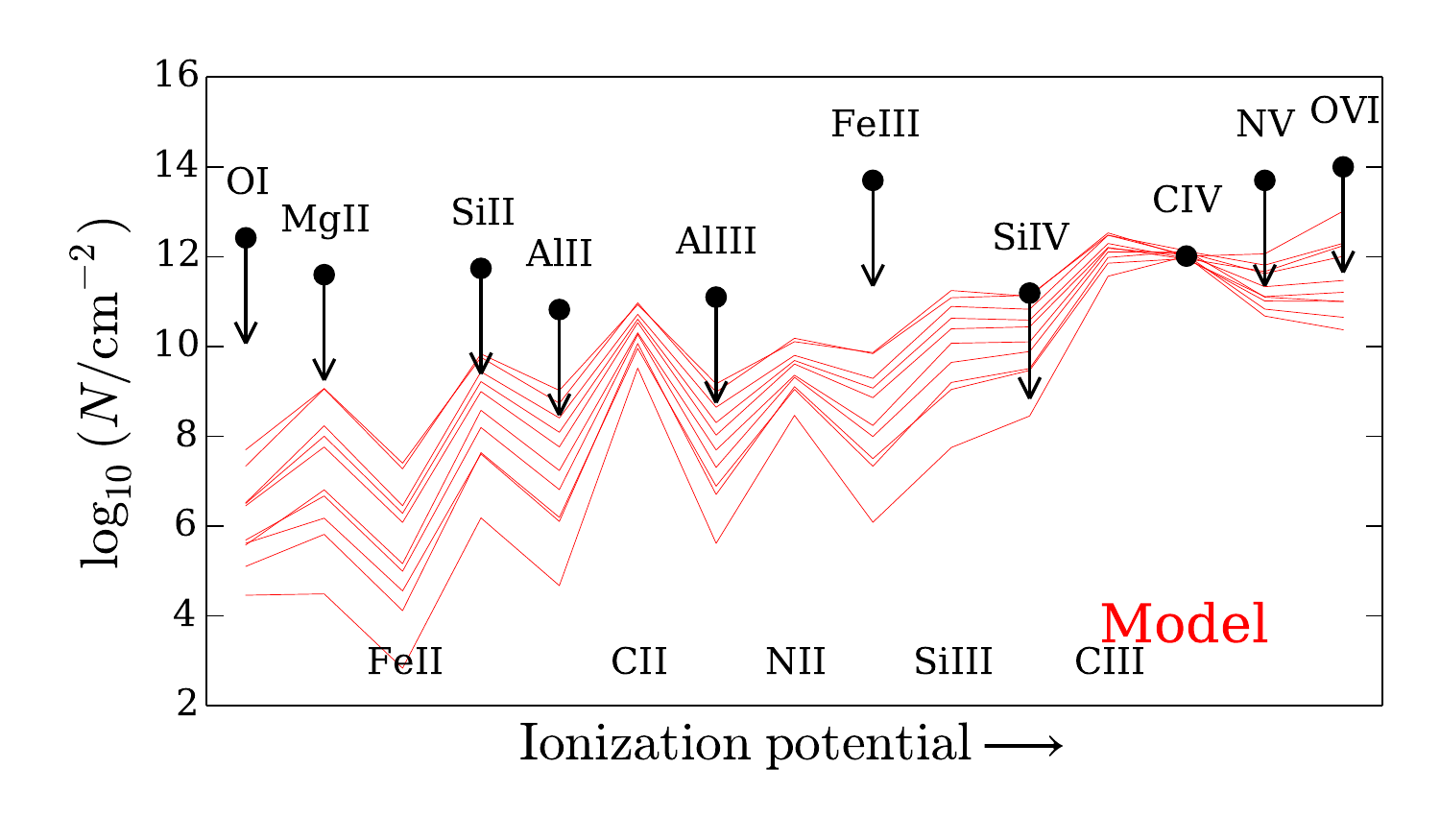}
\caption{\label{f_model8} Ten MCMC samples for component 8. See the Fig.\ref{f_model1} caption for details.}
\end{figure}

After adding \NHI\ and $\alpha_\mathrm{UV}$ the likelihood function becomes
\begin{equation}
\ln \mathcal{L}(Z,\nH,\NHI,\alpha_\mathrm{UV}) = \sum\limits_{i\,=\,1}^\mathcal{N}
\ln(\,\ell_i(Z,\nH,\NHI,\alpha_\mathrm{UV})\,)
\end{equation}
\noindent Therefore, we generate a 4-D grid of \textsc{cloudy} models
as a function of $Z$, \nH, \NHI\ and $\alpha_\mathrm{UV}$. We assume a
plane parallel slab, solar abundance ratios, no dust and
photoionization equilibrium.  To find the posterior distributions for
$Z$, \nH, \NHI\ and $\alpha_\mathrm{UV}$ given this likelihood
function and any priors on the estimated parameters, we use the MCMC
sampler \textsc{emcee} \citep{ForemanMackey13}. This generates the
plots shown in Figures \ref{f_pos1} and
\ref{f_pos2}--\ref{f_pos8}. Figures \ref{f_model1} and
\ref{f_model2}--\ref{f_model8} show ten randomly-selected parameter
samples from the Markov chain sampling compared to the observed column
densities in each component.

For observed column density measurements, we assume the likelihood is
a Gaussian with two discrete $\sigma$ width parameters, one each for
the upper and lower uncertainties. For upper limits we assume a
step-like function with a one-sided Gaussian of fixed $\sigma= 0.05$
dex describing the likelihood above the upper limit. We use an
analogous function for lower limits.  When calculating the likelihood
of a set of model parameters we apply a minimum uncertainty in
log$_{10}$ of the observed column densities of 0.1 dex. This allows
for the fact that there could be (a) more complicated shapes for the
UV background than our single-parameter model can capture, (b) minor
abundance variations, (c) non-equilibrium effects, (d) geometric
effects or (e) density variations across the cloud, which would all
perturb the column densities from the equilibrium, constant density,
solar-abundance models that \textsc{cloudy} calculates. It also
prevents one or two observed values with very small uncertainties from
dominating the final solution. Applying a minimum uncertainty of 0.15
or 0.05 does not significantly change our results.

For all the components we analyse here, \NHI\ can be precisely
measured. Therefore we apply a Gaussian prior to \NHI\ given by the
measurement and uncertainty on \NHI. For components 2, 7 and 8, where
$\alpha_\mathrm{UV}$ is poorly constrained, we also apply a Gaussian
prior to $\alpha_\mathrm{UV}$. Limited ranges in $\log_{10}
(Z/Z_\odot)$ ($-3$ to $0.5$), $\log_{10}\nH$ ($-4.5$ to $-1$),
$\log_{10}\NHI$ ($13$ to $17.5$), and $\alpha_{\rm UV}$ ($-2.5$ to
$1$) are also imposed, which act as another effective prior.

As a test that our method can reproduce the results of previous
\textsc{cloudy} analyses, we applied it to components in
\citet{Crighton13_cma} and \citet{Simcoe06}, applying a strong prior
on $\alpha_\mathrm{UV}$ equivalent to fixing the incident radiation
field shape. We verified that the densities and metallicities we infer
are consistent with the values found by those analyses, given the
ionizing spectral shapes they assumed (HM12 or starburst-dominated for
Crighton et al., and a Haardt-Madau background combined with starburst
spectrum for Simcoe et al.).

\subsection{Normalisation of the incident radiation field and \HeII\ constraints }
\label{a_norm}

The comparison between observed column densities and \textsc{cloudy}
models usually gives tight constraints on the ionization parameter,
$U$. These can be related to the density \nH\ using equation
\ref{e_U}. However, a higher \nH\ value can always be compensated by a
higher normalisation for the incident radiation field, leaving $U$
unchanged. Therefore there is an additional uncertainty in \nH\ due to
the uncertainty in the normalisation of the incident radiation
field. 

For a given value of $\alpha_\mathrm{UV}$ we normalise the ionizing
spectrum such that the photoionization rate, $\Gamma_{\rm HI}$,
matches the value measured by \citet{FaucherGiguere08_gamma} for the
integrated UV background from QSOs and galaxies at $z=2.5$. The
uncertainty in this depends not only on measurement errors, but also
on assumptions about the thermal history of the IGM.  For example,
\citet{Becker13_gamma} find a $\Gamma_{\rm HI}$ $\sim0.3$ dex higher
than Faucher-Giguere et al. by assuming a different thermal
history. To take into account this uncertainty, we adopt an additional
uncertainty of $0.3$ dex for our \nH\ estimates, which we add in
quadrature to the uncertainties estimated from our MCMC procedure. It
is also possible that the true radiation field could be much stronger
than this if there are local ionizing sources from an AGN or
starburst. This is unlikely for the absorber in this work, because the
nearby galaxy has a modest SFR and shows no evidence for an AGN, but
it remains a possibility. It implies that the normalisation we assume
for the radiation field is a lower limit, as local sources can only
add to the background radiation field. Therefore our $\nH$
measurements are lower limits, and our inferred cloud sizes and masses
are upper limits. Local sources can also change the shape of the
ionizing spectrum, as shown in Fig.~\ref{f_aUV1b}. However, the range
of $\alpha_\mathrm{UV}$ we marginalise over covers the expected slopes
from dusty starburst-dominated to QSO-dominated shapes over the 1--10
Ryd range, so while there could be small-scale variations in the shape
of the radiation field due to local sources which our slope
parametrisation does not capture, we do not expect this to strongly
bias the derived $\nH$.

We can also constrain the normalisation of the spectrum above $4$~Ryd
by using constraints on the photoionization rate of \HeII,
$\Gamma_{\rm HeII}$. Using a reference HM12 spectrum at $z=2.5$,
$\Gamma_{\rm HeII}$ ranges from $0.05\times10^{-14}$~s$^{-1}$ for
$\alpha_{\rm UV}=-2$ to $3\times10^{-14}$~s$^{-1}$ for $\alpha_{\rm
  UV}=1$. The most recent constraints on the ratio of \HeII\ to
\HI\ volume densities, $\eta$, are $30< \eta < 200$ at $z\sim2.5$
\citep{McQuinn14}. We can estimate $\Gamma_{\rm HeII}$ from $\eta$ and
$\Gamma_{\rm HI}$ using equation 18 in McQuinn \& Worseck:
\begin{equation}
\Gamma_{\rm HeII} \approx 0.43\,\Gamma_{\rm HI} /\eta.
\end{equation}
Assuming $\Gamma_{\rm HI} \sim 0.8 \times 10^{-12}$~s$^{-1}$
\citep{FaucherGiguere08_gamma} implies $0.2<\Gamma_{\rm
  HeII}/(10^{14}~\mathrm{s^{-1}})<1.2$. Therefore the range of
$\alpha_\mathrm{UV}$ consistent with $\Gamma_\mathrm{HeII}$
constraints is $-1\lesssim\alpha_\mathrm{UV}\lesssim 0.5$. All our
measured $\alpha_\mathrm{UV}$ values are consistent with this range.

Since we normalise the incident radiation for each value of
$\alpha_{\rm UV}$ independently, the conversion multiplier between
\nH\ and $U$ is not the same for different ionizing spectrum
shapes. We use spline interpolation to find the \nH\ to $U$ conversion
factors when calculating $U$ from our \nH\ parameter samples.

\end{document}